\documentclass[aps,amsmath,amssymb,notitlepage,noeprint,nofootinbib,superscriptaddress]{revtex4-1}
\pdfoutput=1
\usepackage[latin9]{inputenc}
\usepackage{braket}
\usepackage{xcolor}
\usepackage{verbatim}
\usepackage{graphicx}
\usepackage{esint}
\usepackage{epstopdf}
\usepackage[colorlinks,linkcolor=purple,citecolor=green!50!black,urlcolor=blue!70!black]{hyperref}
\hypersetup{pdfauthor={Facoetti et al.},pdftitle={Classical glasses, black holes, and strange quantum liquids}}

\DeclareMathOperator{\re}{Re}
\newcommand{\e}{\textrm{e}}
\newcommand{\di}{\mathrm d}
\newcommand{\imi}{\mathrm{i}}
\newcommand{\ethr}{\mathcal{E}_{\textrm{th}}}
\newcommand{\ediff}{\ethr-\mathcal{E}}
\newcommand{\emin}{\mathcal{E}_{\textrm{min}}}
\newcommand{\Hsyk}{H_{\mathrm{SYK}}}
\newcommand{\Teff}{T_{\mathrm{eff}}}
\newcommand{\Hfp}{H_{\mathrm{FP}}}
\DeclareMathOperator{\tr}{Tr}
\DeclareMathOperator{\Dtwo}{D_1^{(2)}}
\DeclareMathOperator{\sign}{sgn}
\newcommand{\kp}{k}

\newcommand{\ie}{i.e.}

\newcommand{\LPENS}{Laboratoire de Physique de l'Ecole normale sup\'erieure, ENS, Universit\'e PSL, CNRS, 
Sorbonne Universit\'e, Universit\'e de Paris, F-75005 Paris, France}
\newcommand{\ColumbiaChem}{Department of Chemistry, Columbia University, New York, New York 10027, USA}
\newcommand{\KCLMaths}{Department of Mathematics, King's College London, Strand, London, WC2R 2LS,
United Kingdom}

\begin{document}

\title{Classical glasses, black holes, and strange quantum liquids}

\author{Davide Facoetti}
\email{dfacoet@gmail.com}
\affiliation{\LPENS}
\affiliation{\ColumbiaChem}
\affiliation{\KCLMaths}
\author{Giulio Biroli} 
\affiliation{\LPENS}
\author{Jorge Kurchan}  
\affiliation{\LPENS}
\author{David R. Reichman}
\affiliation{\ColumbiaChem}

\begin{abstract}
From the dynamics of a broad class of classical mean-field glass models one may obtain a quantum model with finite zero-temperature entropy, a quantum transition at zero temperature, and a time-reparametrization (quasi-)invariance in the dynamical equations for correlations. The low eigenvalue spectrum of the resulting quantum model is directly related to the structure and exploration of metastable states in the landscape of the original classical glass model.  This mapping reveals deep connections between classical glasses and the properties of SYK-like models.
\end{abstract}

\maketitle

\section{Introduction}
\label{sec:intro}

Recently, there has been an intense activity focused on the Sachdev--Ye--Kitaev (SYK) model~\cite{SachdevYe1993,KitaevLectures}, triggered by the realization that it saturates a quantum bound on the Lyapunov exponent~\cite{Maldacena2016bound}, has non-zero entropy in the limit of zero temperature (taken after the large-$N$ limit) and a temperature-linear specific heat,  just as expected from simple models of black holes~\cite{Sekino2008}.  At the core of the analogy is the fact that the SYK model has an (almost) soft mode with respect to time reparametrizations, a fact that is true at low temperatures in the infrared, low frequency limit. A near-invariance suggests the construction of a ``sigma model'' describing  the system in terms of the cost of reparametrizations. Such a description, in terms of a ``Schwarzian action,'' has been constructed~\cite{Maldacena2016SYK}, providing the ``gravity'' counterpart of the fermionic system, so that the SYK model becomes a toy model of holography. 

In this work we investigate the relationship between the SYK model and classical glassy physics. A formal connection appears already at the level of the Hamiltonian, where the SYK model provides a fermionic analog of the classical $p$-spin model which plays an important role in the physics of the glass transition~\cite{Gardner1985,Kirkpatrick1987}. 

A more physical and direct connection was pointed out by Parcollet, Georges and Sachdev~\cite{Parcollet1999,Georges2000,Georges2001} who studied the quantum Heisenberg spin glass, from which the SYK model emerges as an effective theory.  They showed that the critical behavior captured by the SYK model is actually related to a spin-glass (or glass) transition at zero temperature. By taking 
a rather different path, in what follows we show that this analogy can be pushed much further,  establishing a strong relationship between SYK behavior and {\em classical glass  dynamics}. 
Remarkable facts taking place in the SYK model, such as the existence of a finite zero-temperature entropy, a non-trivial temperature dependence of the specific heat, critical behavior, and an
approximate reparametrization invariance, all find natural counterparts within the picture resulting from the manner in which classical glass physics emerges. In addition, since the same kind of time-reparametrization quasi-invariance which exists in the SYK model also appears in models of glassy dynamics, the relationships we expose provide very fruitful tools for addressing major problems in glassy dynamics.  

\subsection{Main results}

Our main idea is to establish an analogy between SYK and glass physics not directly based on the free energy of the respective models,  but rather through the mapping between stochastic dynamics and a quantum Hamiltonian.  Such a correspondence has been used already several times in the past in condensed matter physics (notably by  Rokhsar and Kivelson~\cite{RokhsarKivelson}), in quantum field theory (for example in stochastic quantization), and in statistical in physics~\cite{ParisiBook}.
The classical-quantum mapping has also been used to construct quantum Hamiltonians from the Fokker--Planck operator associated with classical glasses~\cite{BiroliChamonZamponi,Nussinov2013,Chen2015,Lan2018}, as is done in this work.

\begin{itemize}

\item \textbf{Strange quantum liquid.} \\
Following this mapping, we  shall consider quantum Hamiltonians obtained from the Fokker--Planck operators associated with the classical Langevin dynamics of mean-field glassy systems. 
The eigenvalues of the Fokker--Planck and its associated Schr\"odinger-like operators are in one-to-one relation to the metastable states of the original diffusive model, the eigenvalue of the former with the inverse lifetime of the latter. Moreover, if one considers the sum over periodic trajectories of period $\beta_q=1/T_q$ of this dynamics, one obtains a partition function of the quantum form, whose value equals the number of metastable states of lifetime $\beta_q$ of the diffusive system. The resulting quantum model displays at low temperature a series of remarkable properties, that we can connect precisely to those 
of glassy dynamics. For instance, the resulting quantum models have a non-zero entropy at zero temperature, which is directly related to the large number of metastable states of the parent classical glassy system. They have a critical behavior approaching zero temperature that is linked to the critical properties of glassy dynamics. 

    \item \textbf{Time-reparametrization invariance.}\\
    Time-reparametrization (quasi-)invariance 
    was  initially encountered 
 in the earliest studies of spin-glass dynamics~\cite{Sompolinksy1982} and, implicitly, in the mean-field dynamical framework of glassy behavior known as mode-coupling theory (MCT)~\cite{GotzeBook,GotzeLesHouches,Reichman2005}. Later, when the out-of-equilibrium dynamics of model mean-field glasses was analytically treated~\cite{CugliandoloKurchan},  an exact solution was found {\em up to reparametrizations} with the precise matching of solutions left undetermined.
Apart from this inconvenient matching problem, the question of the physical meaning of  time-reparametrization invariance in the glassy context arose. Physically, the soft mode in the dynamics of a system near or below the glass transition is related to the correlated motion of larger and larger clusters of particles, a process called  {\em dynamical heterogeneity}~\cite{Berthier2011dynamicalHet,BerthierBiroli2011RMP}.  The divergence of the length scale associated with dynamical heterogeneity at the (dynamical) glass transition is quantified by the divergence of a particular four-point function called {$\chi_{4}(t)$}~\cite{Franz2000nonlinear}, which also diverges at zero temperature in the SYK model~\cite{Maldacena2016SYK}.  At the same time, the system develops a growing susceptibility towards certain perturbations, such as shear, which have the effect of dramatically 
reparametrizing the time-dependence of correlation and response functions. 
This phenomenon was actually used to probe correlated motion in experiments via non-linear responses~\cite{Albert2016}. 
In a series of papers~\cite{Castillo2002,Castillo2003,Chamon2002,Chamon2004,Chamon2007,Chamon2011} it was emphasized that reparametrization invariance is a central fact of glassy dynamics, and  a detailed investigation of realistic glass models was performed, culminating in the proposal of an expression for an action playing the role of the Schwarzian theory in the SYK model. It
takes into account the spatial, but not temporal, dependence of reparametrizations, see Eq.~(30) of Ref.~\cite{Chamon2007}. The strange quantum liquid obtained through the mapping with classical glassy dynamics displays time-reparametrization (quasi-)invariance just as the SYK model does. This fact allows us to bridge the gap between these two different incarnations of time-reparametrization invariance and offers a 
promising route to follow to develop a full theory of dynamical fluctuations in glasses. 

\end{itemize}

The theoretical analysis we develop in this work shows that, all in all, glassy dynamics leads us to a (non-fermionic) {\em quantum} model of what we call a ``strange quantum liquid'' with finite entropy in the low-temperature limit, a critical (gapless) point at zero temperature, time-reparametrization quasi-invariance, and possible  quantum effects related to chaotic scrambling. To this extent, some of the remarkable properties of the SYK and related models appear to be already embedded in the manner in which mean-field classical glasses explore their energy landscape. 

\section{Models and Quantum to Stochastic Dynamics Mapping}
\label{sec:models}

The purpose of this section is to introduce the models that are central to this work and the mapping from stochastic to quantum dynamics that we will use to relate the SYK model to classical glassy physics. This section provides background and sets the stage for the following analysis.  
\subsection{The Sachdev--Ye--Kitaev model}
Sachdev and Ye~\cite{SachdevYe1993} introduced a disordered fermionic model which becomes gapless at $T=0$, providing an explicitly solvable model of a quantum Heisenberg spin glass. Later, Parcollet, Georges and Sachdev~\cite{Parcollet1999,Georges2000,Georges2001} studied more
general spin representations leading both to fermionic and bosonic models. They made the observation that low temperature properties of such models have analogies to those of a conformal theory, and identified time reparametrization as the origin of this coincidence.
The situation captured the attention of a larger community when Kitaev~\cite{KitaevLectures} discovered that indeed soft reparametrization modes are responsible for the system generating a behavior that mimics that of a ``toy'' model of a black hole.  In particular a low-temperature dynamics that saturates a quantum bound on chaotic scrambling~\cite{Maldacena2016bound}. He did this employing a slightly simplified variant of the Sachdev--Ye model, with Majorana rather than complex fermions, which makes calculations easier. 

The Hamiltonian of the Sachdev--Ye--Kitaev model reads
\begin{equation}
\Hsyk =  {(\imi)^{\frac{q}{2}}} \sum_{1\leq i_1<\dots<i_q\leq N} J_{i_1...i_q} \chi_{i_1} ...  \chi_{i_q}~,
\label{eq:SYK-H}
\end{equation}
where $\chi_i$ are $N$ Majorana fermions. The couplings $J_{i_1...i_q}$ are independent, identically distributed Gaussian random variables with zero mean and variance $N^{1-q} J^2 (q-1)!$.

In order to study the thermodynamics, one has to compute
\begin{equation}
\overline {\ln  Z}= \overline{\ln  \tr \left[  e^{-\beta \Hsyk} \right]}~,
\end{equation} 
where the overbar denotes an average over the couplings. This is usually done by replicating the system $n$ times and then continuing to $n\rightarrow 0$.
It turns out, however, that due to the Grassmannian nature of the degrees of freedom and the lack of a glass transition, order parameters coupling different replicas vanish, and the result (at least, to leading order in $N$) coincides with the annealed average,
\begin{equation}
 \ln \overline Z= \ln \overline{ \tr \left[ e^{-\beta \Hsyk} \right]}~.
 \label{annealed}
\end{equation} 
The partition function~\eqref{annealed} may be expressed as a path integral, and after averaging over the $J$'s, all fermionic degrees of freedom may be integrated out, resulting in an action purely in terms of the correlation function
\begin{equation} \label{GG}
G(t) =  \sum_i \langle T \chi_i (t) \chi_i (0)\rangle~.
\end{equation}
Thus, one obtains
\begin{equation} 
 \ln \overline{Z}= \ln \int D[G] e^{-N S[G]}~,
\end{equation}
with 
\begin{equation}
 S[G]= \frac{1}{2}\int dt dt'\left\{ \frac{\partial G(t,t')}{ \partial t} - \frac{J^2}{q} G(t,t')^q \right\} + \frac{1}{2} \tr \ln G,
 \label{eq:SYK-action}
\end{equation}
where the logarithm and the trace are for $G$ considered as an operator with convolutions. 
The large $N$ limit allows for a saddle point evaluation, and one finds, after convolving with $G$, that the saddle-point value of $G$ satisfies the following equation
\begin{equation} \label{eq:SYK-eom}
{ \frac{\partial G(t_1,t_2)}{\partial{t_1}}} - J^2 \int_0^\beta d t\ G(t_1, t) \ G(t, t_2)^{q-1} =  \delta(t_1 - t_2)~.
\end{equation}
The same result can be obtained by a consideration of the $N$-dependence of diagrams in the expansion of the self-energy for which only melonic terms remain at large $N$.

The analysis of these equations have revealed three main properties~\cite{Parcollet1999,KitaevLectures,Maldacena2016SYK,Polchinski2016,Jevicki2016}:

\begin{itemize}
\item \textbf{Zero-temperature criticality:} The large $t$ solution of~\eqref{eq:SYK-eom}, for $T \rightarrow 0$, is found to be
\begin{equation} \label{eq:SYK-G0}
G(t_1,t_2) \sim \frac{b}{|t_1-t_2|^{{2}/{q}}}\sign(t_1-t_2),
\end{equation}
where $b$ is a constant depending on $J$ and $q$~\cite{Maldacena2016SYK}.
This critical power law behavior is cut off at small temperature on a timescale of the order of $\beta$. 
\item \textbf{Non-standard thermodynamics:} The specific heat is linear at low temperature and the model displays a positive zero-temperature entropy. 
\item \textbf{Reparametrization (quasi-)invariance:} Equation~\eqref{eq:SYK-eom} has, to the extent that we may neglect the time-derivative term, the approximate reparametrization invariance:
\begin{equation}
G(t_1, t_2) \rightarrow  |\dot h(t_1) \dot h (t_2)|^{1/q} G(h(t_1), h(t_2))~.
\label{repa-syk}
\end{equation}
Substituting $\tilde G(t_1,t_2) =\int_{t_2}^{t_1}  dt \int_{t_2}^{t}  \; dt' \; G^q(t,t') $ into the above form yields
$
\tilde G(t_1, t_2) \rightarrow \tilde  G(h(t_1), h(t_2))~.
$

In reality, only one specific parametrization corresponds to the true minimum of the action.

As noted by Parcollet and Georges~\cite{Parcollet1999}, in analogy with the case of conformal field theories~\cite{TsvelikBook},  the
reparametrization
 $t_a \rightarrow  \tan \left(\frac{\pi t_a}{\beta}\right)$
 ($a=1,2$)  maps~\eqref{eq:SYK-G0} into a {\em time-translational invariant}  function of period $\beta$,
\begin{equation}
G_{\beta}(t_1-t_2) = b \left[\frac{\pi}{\beta \sin\frac{\pi(t_1-t_2)}{\beta}}\right]^{2/q} \sign(t_1-t_2),
\label{eq:SYK-Gbeta}
\end{equation}
so that the low-temperature behavior is obtained by reparametrization of the zero-temperature one.

The breaking of reparametrization invariance, which is a continuous symmetry, leads to the emergence of almost-soft modes governing low temperature fluctuations. 
An effective theory based on it allows one to compute the main critical fluctuations, which corresponds to four-point functions, 
in particular those related to the quantum Lyapunov exponent---extracted from the so called out-of-time-order correlation function (OTOC). 
\end{itemize}
As we shall show, the relationship with glassy physics presented in this work will give a context where there is a natural 
interpretation of the first two points and unveil promising connections for the third. 

\subsection{The \texorpdfstring{$p$}{p}-spin spherical model}
The $p$-spin spherical model was introduced in Ref.~\cite{Crisanti1992},
  \begin{equation}
E=\sum_{i_1<...<i_p} J_{i_1,...,i_p} q_{i_1}... q_{i_p}  \;\;\;\;\;\;\;\; , \;\;\;\; \;\;\;\;\sum_i q_i^2=N~,
\label{spherical}
\end{equation}
where $q_i$ are $N$ real-valued ``soft-spins'' obeying the spherical constraint $\sum_i q_i^2=N$, which replace the binary $s_i=\pm 1$ Ising spins. The couplings are random variables as in the SYK model~\eqref{eq:SYK-H} (couplings with repeated indices are set to zero). The model is a generalized spin-glass model introduced in the early days of spin-glass theory and that later played a central role 
in the theory of the structural glass transition, as we shall briefly recount in the next section for completeness.  Its classical~\cite{Crisanti1992} and quantum-mechanical~\cite{Cugliandolo2000,Cugliandolo2001} thermodynamics has been studied by the replica method. 
Here, although we attempt a connection with the (quantum) SYK model, we shall only need to restrict ourselves to the {\em dynamics} of the {\em classical} glass  mimicking the interaction with a thermal bath of temperature $T$. This stochastic dynamics, based on the Langevin equation, can be analyzed using field theoretical methods such as Martin--Siggia--Rose--Janssen--De Dominicis (MSRJD) formalism, a path-integral approach for the evolution of the probability
distribution~\cite{MartinSiggiaRose1973,Janssen1976,DeDominicis1976}, see Refs.~\cite{Kurchan09-lectures,Hertz2016MSRJD,Zinn-Justin_Book} for introductions to the MSRJD construction.
The equilibration time diverges with $N$ at a temperature $T_d$, the ``dynamical'' transition temperature, below this temperature the equilibration
time becomes infinite and we may study the slow (unsuccessful) approach to equilibrium.

\begin{figure}
 \includegraphics[width=0.4\columnwidth]{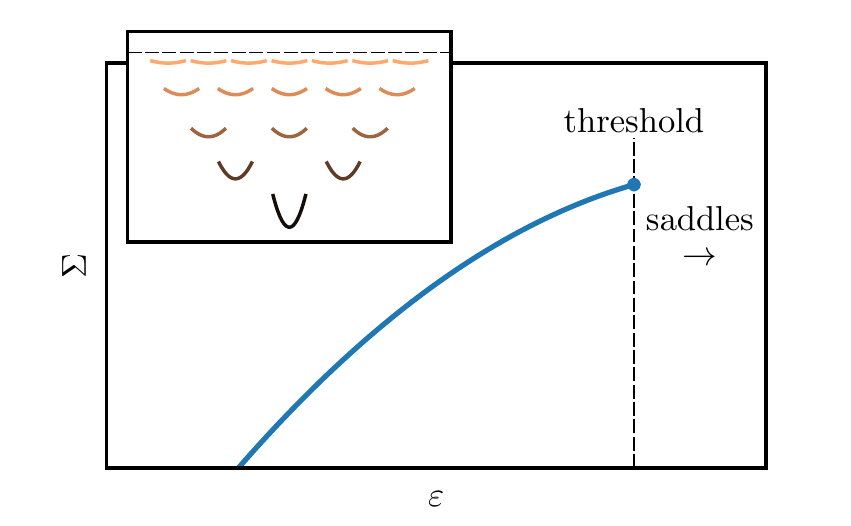}
 \caption{\label{fig:complexity}Complexity $\Sigma$ of the energy landscape (see Eq.~\ref{eq:complexity}). Below a threshold energy density $\ethr$ the number of minima
 is exponential in $N$. Above the threshold the critical points are saddles.
 Inset: deeper minima have higher curvature, and become marginal approaching the threshold.}
\end{figure}

Using the MSRJD path integral approach, one can follow a procedure very similar to the one sketched in the previous section for the SYK model.  First, one obtains a field theory for the two-point functions, which can then be solved in the large $N$ limit by the saddle-point method. One finds an equation which, in the high-temperature phase $T>T_d$ reads,
in terms of the correlation function $C(t-t')=\frac 1 N \sum_i \langle q_i(t) q_i(t')\rangle$,
  \begin{equation}
\frac{dC(t_1-t_2)}{dt_1}=-TC(t_1-t_2)-\frac{p}{2T}\int_{t_2}^{t_1} dt C^{p-1}(t_1-t)\frac{dC(t-t_2)}{dt}~,
\label{MCT}
\end{equation}
where $\langle \cdot \rangle$ means the average over the thermal noise. We postpone for a later section the discussion of what becomes of this equation below $T_d$.
This ``mode-coupling'' equation~\eqref{MCT} shows a striking similarity with the corresponding equation of motion for the Green's function of the SYK model~\eqref{eq:SYK-eom}.  It should be noted that a diagrammatic approach (which, in the case $p=4$, selects only the ``melonic'' terms in the large $N$ limit) also leads to the exact equation~\eqref{MCT}, see Ref.~\cite{Bouchaud1996MCT}. A large body of work has shown that the $p$-spin spherical model displays three main properties: 

\begin{itemize}
\item \textbf{Dynamical criticality:} When the temperature approaches $T_d$ the solution of~\eqref{MCT} shows a two-step behavior: the correlation first decays to a plateau value, $q_{EA}$, and then departs from it and decreases to zero. The timescales for these two decays both diverge approaching $T_d$ as power laws, the latter as $\tau_\alpha(\epsilon)\sim\epsilon^{-1/2a-1/2b}$ and the former as $\tau_\beta\sim \epsilon^{-1/2a}$, where $a,b$ are positive exponents ($\epsilon=T-T_d$). The approach and the departure from the plateau value follow  power laws
$$
C(\tau)=q_{EA}+\frac{c}{\tau^a} \qquad 1\ll\tau\ll \tau_\beta\,,\qquad C(\tau)=q_{EA}-c'\left(\frac{\tau}{\tau_\alpha}\right)^b
\qquad \tau_\beta\ll\tau\ll \tau_\alpha\,.
$$
A similar dynamical criticality exists also in the out-of-equilibrium dynamics induced by quenches below $T_d$.

\item \textbf{Time reparametrization (quasi-)invariance:}
It is relevant here to recall how these exponents are obtained~\cite{GotzeBook,GotzeLesHouches}.  Close to $T_d$ one concentrates on the long-time (infrared) behavior
and makes use of i) a Taylor expansion of $C(t)-q_{EA}$ and ii) neglects the time derivative in~\eqref{MCT}, thus obtaining time reparametrization invariance.
Combining a uniform time stretching with a rescaling of the ``field'' $ C(t) -q_{EA}$, one obtains a form  $C(t) -q_{EA}= \epsilon^{1/2}  g\left(\epsilon^{2a} t \right)$, which works with a single $g(t)$ for all $T$ just above $T_d$, i.e. for small $\epsilon$.  The existence of a scaling form with a universal $g$ is referred to as the ``time-temperature superposition principle'' and is a consequence of reparametrization invariance~\cite{GotzeBook}.

The equations for the aging dynamics also display, everywhere below $T_d$, time-reparametrization invariance at long times, such that the time derivative may be neglected. 
Time-reparametrization invariance is only exact in the zero frequency limit, but remains as a generator of a soft mode governing long-time dynamical fluctuations~\cite{Chamon2007}, in particular the ones known as dynamical heterogeneities that are probed by four-point correlation functions such as what is referred to as $\chi_{4}(t)$ in the glass literature~\cite{Kob1997,Berthier2007}.   At $T_d$ $\chi_{4}(t)$ diverges precisely as it does in the SYK model at $T=0$ and for precisely the same reasons~\cite{Maldacena2016SYK,Kirkpatrick1988,Biroli2004}.

\item \textbf{Complex energy landscape:} The energy of the $p$-spin model~\eqref{spherical} has a number of minima which is exponential in $N$, see Fig.~\ref{fig:complexity}.  There is a range of energies between the minimum $\emin$ and a threshold value $\ethr$ where minima exist; for energies higher than $\ethr$ there are only saddles~\cite{KurchanLaloux1996}. Both quantities are ``self averaging,'' meaning that their deviation from one realization of $J$'s to another vanishes in the thermodynamic limit.
The number of minima grows exponentially with the energy, 
\begin{equation}
 N(E) = e^{N \Sigma(E/N)} \; , \;\;\;\;\;  1/\Teff(E) \equiv  N \frac{d  \Sigma}{dE}~.
 \label{eq:complexity}
 \end{equation}
 The function $\Sigma(\mathcal{E})$ is known as the ``complexity,''
and vanishes abruptly at the threshold (see Fig.~\ref{fig:complexity}). 
Its derivative defines the effective temperature $\Teff$.
The exponential dependence of the number of minima 
 implies that the vast majority of minima lie just below the threshold.
\end{itemize} 
In conclusion, not only do the classical $p$-spin spherical spin glass and the SYK models display very similar Hamiltonians, but in addition the stochastic dynamics of the former shows enticing similarities with the low temperature imaginary time quantum dynamics of the latter. Yet, it is not clear how to go beyond these analogies. Constructing a strong connection 
between the SYK model and the classical glassy behavior of~\eqref{spherical} is precisely the purpose of the remainder of our work. We shall accomplish this by exploiting a mapping from stochastic to imaginary time quantum dynamics that will be reviewed in the next section. 

\subsection{Classical to quantum: from classical glasses to strange quantum liquids}
We now recall a connection between stochastic and quantum dynamics that has been already used several times in the past 
in statistical physics, condensed matter and quantum field theory~\cite{RokhsarKivelson,ParisiBook,Kurchan09-lectures}. 
We consider a system of $N$ coupled degrees of freedom $q_i(t)$ evolving by stochastic Langevin dynamics
	\begin{equation}
	\dot{q_i}(t) = -\frac{\partial V}{\partial q_i} + \eta_i(t)~,
	\end{equation}
	where $V$ is the interaction potential, $T_s$ the (classical) temperature of the thermal bath to which the system is coupled, and $\eta_i(t)$ is a Gaussian white noise with covariance $\langle\eta_i(t)\eta_i(t')\rangle=2 T_s \delta(t-t')$.
	The evolution of the probability density is generated by the
	Fokker--Planck operator $H_{\textrm{FP}}$,
	\begin{equation}
	\partial_t P_t(\mathbf{q}) = \sum_i \frac{\partial}{\partial q_i}
	\left[T_s \frac{\partial}{\partial q_i} + \frac{\partial V}{\partial q_i}\right] P_t(\mathbf{q}) \equiv - H_{\textrm{FP}}P_t(\mathbf{q}) .
	\end{equation}
	
	The Fokker--Planck operator is not Hermitian, but detailed balance is satisfied
	with the Gibbs distribution
	\begin{equation}
	\e^{V/T_s} H_{\textrm{FP}} \e^{-V/T_s}=H_{\textrm{FP}}^\dag~.
	 \end{equation}
	 Detailed balance allows us to write this in an explicitly Hermitian form~\cite{Zinn-Justin_Book,Kurchan09-lectures}. Rescaling time,
	one can define the operator
	\begin{equation}\label{eq:FP-to-H}
	H = \frac{T_s}{2} \e^{V/2T_s} H_{\textrm{FP}} \e^{-V/2T_s} = \sum_i\left[-\frac{T_s^2}{2} \frac{\partial^2}{\partial q_i^2}
	+\frac{1}{8}\left(\frac{\partial V}{\partial q_i}\right)^2 -\frac{T_s}{4}\frac{\partial^2 V}{\partial q_i^2}\right]\ .
	\end{equation}
	$H$ has the form of a Schrodinger operator with $T_s$ playing the role of $\hbar$, unit mass and potential
	\begin{equation}\label{eq:Veff}
	V_{\textrm{eff}} = \frac{1}{8}\left(\frac{\partial V}{\partial q_i}\right)^2 -\frac{T_s}{4}\frac{\partial^2 V}{\partial q_i^2}~.
	\end{equation}
	
	The spectrum of $H_{\textrm{FP}}$ and that of $H$ are the same, up to the rescaling in~\eqref{eq:FP-to-H}, and the 
	eigenvectors are related via the transformation above. 
		
    Let us now recall some general facts about stochastic equations. The spectrum of eigenvalues $\lambda_i$ and eigenvectors $\psi_i$ of $H$  (or $\Hfp$) have a direct relation to metastable states of the original diffusive dynamics (\cite{Gaveau1998,Bovier2002}, see also~\cite{BiroliKurchan01}):
\begin{itemize}
\item The equilibrium state has $\lambda_o=0$ and the corresponding right eigenvector of $\Hfp$ is the Boltzmann distribution associated with the energy function $V$ ($\psi_0$ is the square root of the Boltzmann distribution). 
\item Given a timescale $t^*$, the number of eigenvectors with $\lambda_i<\frac{1}{t^*}$ is the number of metastable states of the diffusive model with lifetime larger than $t^*$. {\em In particular, the eigenvalues $\lambda_i\to 0$ in the thermodynamic limit correspond to metastable states whose lifetime diverges with $N$.}
\item The probability distribution within such metastable states is given by linear combinations of the corresponding $\psi_i$ 
multiplied by $e^{-V/2T_s}$. ``Pure'' metastable states are extremal, in the sense that they are the minimal combinations which are essentially greater or equal to zero everywhere.
\end{itemize}

The partition function of the quantum Hamiltonian $H$ reads
\begin{equation}
    Z(\beta_q) = \tr \e^{-\beta_q H} = \tr \e^{-\frac{1}{2}\beta_q T_s \Hfp}\ .
\end{equation}
It can be represented as a Matsubara imaginary-time path integral.
From the classical stochastic process perspective, the analogous construction is that of a MSRJD path integral~\cite{MartinSiggiaRose1973,Janssen1976,DeDominicis1976}, restricted to trajectories that return to the initial point after a time $t^*=\beta_q T_s/2$.
Indeed such a construction was presented by Biroli and Kurchan~\cite{BiroliKurchan01},	who showed that the resulting object $\mathcal{N}(t^*) = \tr \e^{-t^*\Hfp}=Z(\beta_q)$
counts the number of states of the system that are stable up to a time $t^*$ or longer.
The corresponding contribution $\e^{-t^*\lambda_i}$ is of order one for $t^*\lesssim 1/\lambda_i$, and exponentially small after that. A more precise description is given by the Gaveau--Schulman construction~\cite{Gaveau1998,BiroliKurchan01}.

We thus have introduced a ``quantum'' Hamiltonian $H$, which is associated with a quantum temperature $T_q=1/\beta_q$. The original temperature $T_s$ now plays the role of the quantum parameter, $\hbar$. Likewise, our ``quantum energy'' is associated with the eigenvalues of $H$, which are a measure of the lifetimes of the original classical diffusive system.
	
Finally, one can also establish a relationship between the zero-temperature quantum correlation function (for a diagonal Hermitian operator in configuration space) and the equilibrium stochastic correlation function~\cite{Henley2004,BiroliChamonZamponi}. Defining 
\begin{equation}\label{cqi}
	C_q^I(\tau)=\langle A(\tau)A(0)\rangle~,
\end{equation}
the quantum correlation function in imaginary time defined in the ground state, where $\tau$ is the imaginary time, and $C_c(\tau)=\langle A(\tau)A(0)\rangle$ the classical stochastic correlation function at equilibrium, one has $C_q^I(\tau)=C_c(\tau)$. Moreover, one can write 
\begin{equation}
    C_c(\tau)=C_q^I(\tau)=\int_0^\infty \frac{d\omega}{2\pi}\rho_q(\omega)e^{-\omega t},
\end{equation}		
where $\rho_q(\omega)$ is the so-called spectral density~\cite{MahanBook}. Defining real-time quantum correlation functions in the ground state as $C_q^r(t)=\frac 1 2\langle A(t)A(0)+A(0)A(t)\rangle$, one finds 
\begin{equation}
    C_q^r(t)=\int_{-\infty}^\infty \frac{d\omega}{2\pi}\rho_q(\omega)\cos (\omega t)~.
\end{equation}		
These results establish a correspondence between {\em dynamical} properties of the stochastic dynamics, and equilibrium properties of the quantum model. In particular it relates the distribution of classical relaxation times to the quantum spectral density.
In the following we shall exploit this connection to study the low-temperature properties of the $p$-spin spherical model and relate it to SYK physics, thus unveiling the connection between SYK-like physics and classical glasses {\em when considered from the dynamic point of view}.

\section{A very brief history of mean-field glasses}
\label{sec:history}

The purpose of this section is to provide a sketch of the theory of glasses based on mean-field disordered models emphasizing what is relevant for the connection with SYK physics. Experts can readily jump to the next section.

Not long after the discovery of the spin-glass transition in real spin glasses by Canella and Mydosh~\cite{Cannella1972}, Edwards and Anderson proposed their canonical model on a d-dimensional lattice with nearest-neighbor interactions taking the form~\cite{EA1,EA2}
\begin{equation}
E= \sum_{ij} J_{ij} s_i s_j,  \;\;\;\; s_i=\pm 1~,
\label{EA}
\end{equation}
where the $J_{ij}$ are quenched (non-evolving) random Gaussian variables with zero mean and unit variance. 
A mean-field version of~\eqref{EA} soon followed, introduced by Sherrington and Kirkpatrick (SK)~\cite{SK}. This model is the fully-connected version of~\eqref{EA}, with $J_{ij}$ having a variance $1/\sqrt N$, with $N$ the number of spins. The system has a thermodynamic transition at a temperature $T_c$. The thermodynamics of even this mean-field  model turned out to be highly non-trivial to solve for low temperatures $T<T_c$. The full solution was achieved by Parisi in a series of papers~\cite{MezardParisiVirasoro}. The solution used the replica trick, but has been recently confirmed by rigorous mathematics~\cite{Guerra2003,Talagrand}.   
 
 
A few years later, Derrida introduced the Random Energy Model (REM)~\cite{Derrida1980,Derrida1981}, conceived as a toy version of the (already toy) SK model.
It allows for a complete solution using elementary mathematics. In order to justify the model, Derrida pointed out that it may be obtained as the large-$p$ limit of a spin glass related to the SK model, but with $p$-spin interactions:
 \begin{equation}
E=\sum_{i_1<\dots<i_p} J_{i_1,\dots,i_p} s_{i_1}\dots s_{i_p}  \;\;\;\; , \;\;\;\; s_i=\pm 1~.
\end{equation}
Here, the $J_{i_1...i_p}$ have zero mean and variance $J^{2}p!/(2N^{p-1})$.
The thermodynamics of the  model was later solved by Gross, Kanter, and Sompolinsky~\cite{GKS} and by Gardner~\cite{Gardner1985}.
 
A remarkable breakthrough  came in the late 80's, when Kirkpatrick, Thirumalai and Wolynes (KTW) noted that~\cite{KTW} that the models  with $p>2$ differ substantially from the SK model in that they have a thermodynamic transition at temperature $T_k$ obtained with replicas, but the equilibration time diverges with $N$ at a {\em higher} temperature $T_d>T_k$, the ``dynamical'' transition temperature. KTW then went on to argue that this is exactly what one should expect of a mean-field model of a {\em structural} glass (i.e. made of particles),
 which have a different phenomenology than spin glasses. Their bold intuition has been confirmed by a long series of works laying out the exact $d = \infty$ thermodynamic and dynamics of hard-spheres which displays the same phenomenology as the $p>2$ models~\cite{Maimbourg2016,KPUZ}.
 
 It turns out that for $p>2$, it is much easier to work with a system with continuous variables~\cite{Crisanti1992} as in~\eqref{spherical} where  the spherical constraint $\sum_i q_i^2=N$ replaces the binary $s_i=\pm 1$ of the spins.  This is the model we introduced in the previous section and that has a strong resemblance with the SYK model.

As we have already mentioned, the energy landscape of the $p$-spin spherical model has a number of minima that is exponential in $N$.  The associated entropy function, the complexity (see previous section), is positive in a range of energies between the minimum $\emin$ and a threshold value $\ethr$ and stops abruptly at the threshold, see Fig.~\ref{fig:complexity}. This implies that the vast majority of minima lie just beneath the threshold.

The spectrum of the Hessian of the energy in a minimum depends on its ``depth'' beneath the threshold $\ediff$. It is a semicircle (as in random matrices~\cite{RMT}) but shifted so that the lowest eigenvalue is proportional to $\ediff$.
Hence, the deeper below the threshold the minima lie, the more stable they are. States just beneath the threshold---the vast majority---are marginal and thus the spectrum of their Hessian is gapless.
Consistently, the barriers between states are proportional to the depth beneath the threshold~\cite{Ros2018}.

The interpretation of the dynamical transition at $T_d$ is that the system approaches a temperature at which the threshold states, essentially finite-temperature versions of the energy minima, give the main contribution to the Gibbs measure. The equilibrium dynamics therefore slowly surf over nearly stable states at $T=T_d+\epsilon$. 
For quenches below $T_d$ starting from high temperatures, the system does not equilibrate and  
{\em ages}, again evolving  just above the threshold states.

\begin{figure}
 \includegraphics{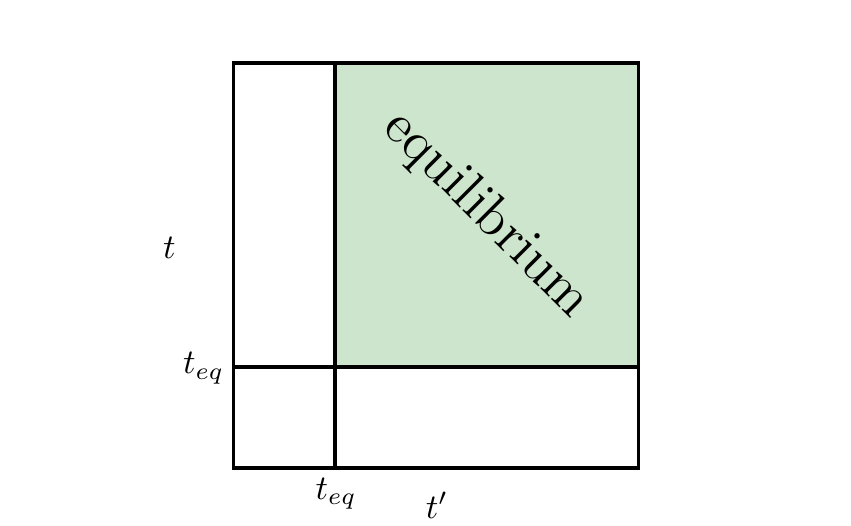}
 \caption{ \label{nonaging} In equilibrating systems correlation functions $C(t,t')$
 become time-translational invariant at long times $t, t'>t_{eq}$.}
\end{figure}

\begin{figure}
 \includegraphics{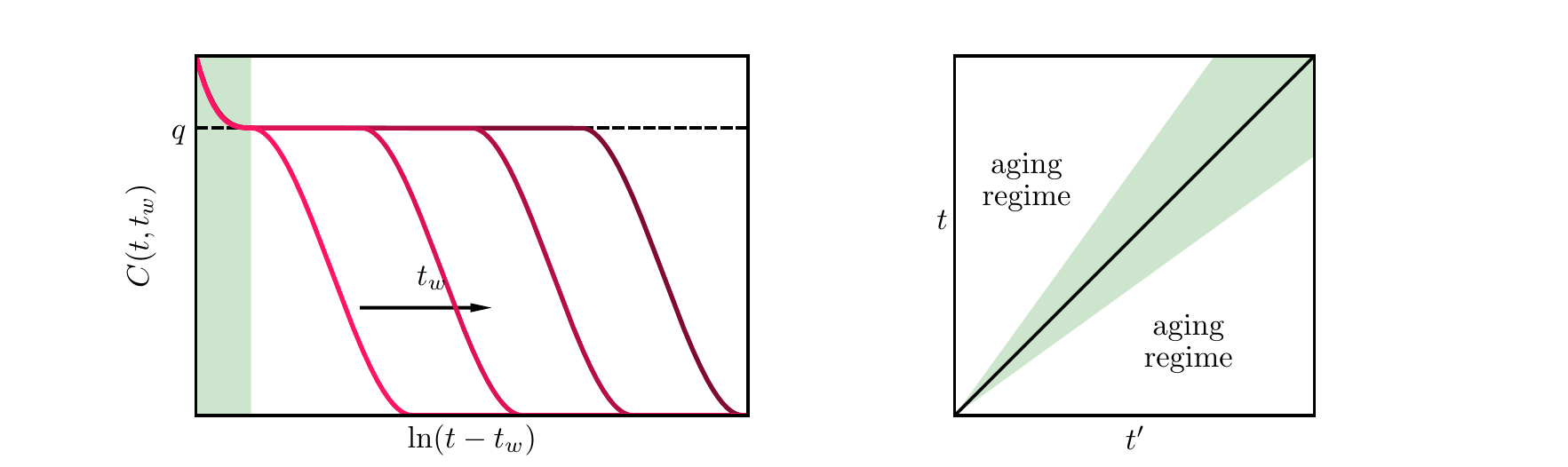}
 \caption{ \label{aging}
Aging phenomenology: quenching from high temperature, there is no time $t_{eq}$ after which the correlation function becomes time-translationally invariant.}
\end{figure}

To understand the difference between these two regimes consider correlations starting from a random (high temperature) configuration and evolving at  $T>T_d$: the situation is depicted in Fig.~\ref{nonaging}. There is a time $t_{eq}$ such that for $(t,t')$ both larger than $t_{eq}$ all two-point functions are {\em stationary}, they depend only on time differences.
If instead we do the same with a bath at $T<T_d$, the outcome is as in Fig.~\ref{aging}. There are two time regimes: when the time-difference is of $O(1)$ the  situation is akin to equilibrium (denoted in green in Fig.~\ref{aging}), while for large times, but  such that $h(t)$ and $h(t')$ are
comparable (for some growing function $h(t)$, for example $h(t)=t$), the correlations scale like ${\cal C}\left(\frac{h(t')}{h(t)}\right)$, a situation manifestly impossible in equilibrium. This situation is called {\em aging} in the glass literature.
The difference in the two regimes emerges in more detail comparing the correlation and response functions,
\begin{equation}
C(t,t')=\frac 1 N \sum_i \langle q_i(t) q_i(t')\rangle \ , \quad R(t,t')=\frac 1 N \sum_i \left.\frac{\partial \langle q_i(t)\rangle}{\partial b_i(t')}\right|_{b_i=0}~,
\end{equation}
where $b_i(t')$ is a field linearly coupled to $q_i$.
\begin{itemize}
\item For $T>T_d$ it takes a finite time $t_{eq}$ for the system to equilibrate, and for  $t\gg t_{eq}$ and $t'\gg t_{eq}$ correlation and response  satisfy the fluctuation-dissipation relation (FDR):
\begin{equation}
T_s R(t-t')= \frac{\partial C(t-t')}{\partial t'}~.
\label{FDT}
\end{equation}
\item For $T<T_d$ as mentioned above, the system {\em never} equilibrates, or rather, it takes a time that diverges with $N$ to do so, and there is no time $t_{eq}$ such that $C$ and $R$ become stationary and satisfy FDR for all
$t\gg t_{eq}$ and $t'\gg t_{eq}$.
In the  high-frequency (ultraviolet) regime where $t-t'$ is finite,
$C$ and $R$ satisfy FDR and are stationary, but in the {\em aging} low-frequency regime with $h(t)$ and $h(t')$ comparable
but arbitrarily large FDR never holds. A  solution in this regime is given by
\begin{equation}
C(t,t')= {\cal C}\left(\frac{h(t')}{h(t)}\right) \;\;\; , \;\;\; R(t,t')= \frac{1}{ \Teff}\dot h(t')  {\cal C'}\left(\frac{h(t')}{h(t)}\right)~.\end{equation}
The constant $\Teff$, the asymptotic energy, and the function ${\cal C}$ have been calculated~\cite{CugliandoloKurchan}. 
The fact that correlation and response functions satisfy an FDR with a (model determined) effective temperature $\Teff$ is
a surprise, see Ref.~\cite{CugliandoloKurchanPeliti}. Note that $\Teff$ here is $\Teff(\ethr)$ of~\eqref{eq:complexity}, a remarkable fact given that the system is not
in equilibrium on the threshold.
\item The function $h(t)$ is well-defined, 
but has not yet been computed analytically. This results from the fact that in this regime there is an approximate
reparametrization symmetry of the problem:
\begin{equation}
\begin{aligned}
C(t,t') &\rightarrow C(h(t),h(t'))~,\\
R(t,t') &\rightarrow \dot h(t') \;R(h(t),h(t'))~,  
\end{aligned}
\label{repa1}
\end{equation}
which becomes more accurate as times become larger, and relaxation slower. Note that in the aging situation the parameter governing reparametrization invariance is $t_w$ and not temperature as in the SYK model. 
\end{itemize}

Both the aging regime and the equilibrium dynamical transition at $T_d$ are dynamical critical phenomena characterized by diverging timescales and correlations. Given that the ``order parameters'' for these transitions are two-point correlation functions, it is natural to expect that critical correlations are encoded in four point functions. This is indeed the case. In particular the fluctuations of the instantaneous correlation function,
\begin{equation}
\chi_4 (t,t') = N \left[ \overline{C^2(t,t')}-\overline{C(t,t')}^2\right]=\frac 1 N \sum_{i,j}\langle s_i(t) s_i(t') s_j(t) s_j(t') \rangle-C(t,t')^2
\end{equation}
have been shown to display critical behavior~\cite{Franz2000nonlinear,Kirkpatrick1988,Biroli2004}. Physically, these fluctuations encode the fact that relaxation is correlated 
from one region to the other of the system, a phenomenon that is observed in experiments and simulations of glassy liquids and goes under the name of dynamical heterogeneity~\cite{Kob1997,Berthier2007}. 

The view we follow here is that, as in the SYK model, critical fluctuations in the four-point functions are due to a soft-mode of the glassy dynamics, which is closely related to time reparametrization (quasi-)invariance. The importance of this soft mode in the context of the aging dynamics  was already 
discussed and tested numerically in Ref.~\cite{Chamon2011}, and we will return to this point at the end of the paper. 

\section{A bridge}
\label{sec:bridge}

The aim of this section is to establish a closer connection between the SYK model and glassy physics. Our starting point is the mapping from 
stochastic to quantum dynamics described in the previous section. 
	
We consider the stochastic Langevin dynamics of the spherical $p$-spin model~\eqref{spherical} in which for simplicity the soft spherical constraint is imposed by a function $f(x)$ with a steep minimum at $x=1$,
	\begin{equation}
	V(q) = -\sum_{i_1<\dots<i_p} J_{i_1\cdots i_p} q_{i_1}\dots q_{i_p} + N f\left(\frac{1}{N}\sum_i q_i^2\right)~.
	\end{equation}
	
	As explained in Sec.~\ref{sec:models}, the Fokker--Planck operator associated to this stochastic dynamics can be mapped into a quantum Hamiltonian $H$ with $T_s$ playing the role of $\hbar$, and a potential
	\begin{equation}
	V_{\textrm{eff}} = \frac{1}{8} \sum_i\left[\sum_{i_2<\dots<i_p}J_{i\, i_2\dots i_p} q_{i_2}\cdots q_{i_p}
	-q_i \lambda\right]^2
	-\frac{T_s}{2}N f'\left(x\right) - T_s x f''\left(x\right)~.
	\end{equation}
	Note that the Laplacian of the $p$-spin term vanishes. Here we have set $x=\frac{1}{N}\sum_i q_i^2$.
	The last term can be neglected since it is subleading for large $N$ and we introduced the definition of the Lagrange multiplier $\lambda=2 f'$.  Our strategy in the following will be to show that $H$ displays an SYK-like physics which can be explained in terms of the glassy properties of the corresponding stochastic dynamics induced by $\Hfp$.

	\subsection{The formalism:  mapping and correlation functions}

Computing the partition function of the quantum problem is equivalent to summing over all periodic trajectories of the stochastic model. This can be done using the MSRJD formalism and proceeding as for the SYK model by the saddle-point 
method~\cite{BiroliKurchan01}. Because trajectories are required to be periodic, causality is broken. At $T_s<T_d$ so is equilibrium, and
we need to consider three, instead of one---as in~\eqref{MCT}---two-point functions:
\begin{equation}
\begin{aligned}
C(t,t') &= \frac 1N \sum_i q_i(t)q_i(t')~,\\
R(t,t') &= \frac 1N \sum_i q_i(t)\eta_i(t')~,\\
D(t,t') &= \frac 1N \sum_i \eta_i(t)\eta_i(t') - 2T_s \delta(t-t')~,
\end{aligned}
\label{CRD}
\end{equation}
where $C(t,t')$ is the correlation function that we have encountered already, and the two other
two-point functions are correlation and response functions that involve the noise history and its correlations with the trajectories.



The two-point functions appearing in the MSRJD formalism are directly related to quantum correlation functions. Calling $\langle \tilde B(t)\tilde A(t')\rangle$ the correlation function obtained from the sum of stochastic periodic trajectories, one has the relation: 
\begin{equation}
\langle \tilde B(t)\tilde A(t')\rangle = \tr\left[  \tilde B e^{-(\beta-t')\Hfp}        \tilde A e^{-t'\Hfp}  \right]/Z=\tr\left[  B e^{-(\beta-t')H}        A e^{-t'H}  \right]/Z~,
\label{trace0}
\end{equation}
where the relation between the operators reads
\begin{equation}
\tilde A= e^{\pm \beta_s V/2} \; A \; e^{\pm \beta_s V/2} \;\;\;\; , \;\;\;\; \tilde B= e^{\pm \beta_s V/2} \; B \; e^{\pm \beta_s V/2}~,
\label{eq:op-change}
\end{equation}
and 
\begin{equation}
Z = \tr\left[   e^{-\beta H} \right]= \tr\left[   e^{-\beta \Hfp} \right]~.
\label{trace1}
\end{equation}
Equation~\eqref{trace0} establishes the connection between classical correlation function within the MSRJD formalism and the quantum ones. 
In practice, it is convenient to work in the original basis of the Fokker--Planck operator, because disorder appears linearly.

 \subsection{The mean-field equations for the periodic trajectories and reparametrization invariance}

Since we consider times of order one with respect to $N$, the functional
 integral for~\eqref{trace1} is dominated by a saddle point contribution. 
We shall obtain periodic dynamic solutions which, in the glassy phase
  (a) break causality, (b) have non-zero action, (c) satisfy 
time-translational invariance, and (d) satisfy time-reversal symmetry.
Defining the expectations $C$, $R$, and $D$ as in~\eqref{CRD} in the Fokker--Planck basis, namely the ``tilde''  operators 
in~\eqref{eq:op-change}, we thus have
\begin{equation}
C(t,t')=C(|t-t'|) \;\;\; , \;\;\; D(t,t')=D(|t-t'|)  \;\;\; \mbox{and}  \;\;\;  T[R(t-t')-R(t'-t)]= \frac{\partial C(t-t')}{\partial t'}~.
\end{equation}
Note that (a) and (b) are properties typical
of instantons, while (c) and (d) are not.
In the high-temperature phase there is a periodic solution with zero action for long times corresponding essentially to the equilibrium dynamics.

By averaging over the disorder and assuming a diagonal replica symmetric ansatz, as done for the SYK model, one obtains a functional integral over $C,R,D$ and a weight of the form $e^{- N S[C,R,D]}$ with
 the action
 \begin{equation}\label{action}
 \begin{aligned}
S&=-\int_{0}^{t^{*}}dt\left.\left(\partial_{t}R(t,t')
+\lambda R(t,t')-TD(t,t')\right)\right|_{t'=t^+}\\
&+\frac{p}{4}
\int_{0}^{t^{*}}dtdt'\left(D(t,t')C^{p-1}(t,t')+(p-1)R(t,t')R(t',t)
C^{p-2}(t,t')\right)\\
&-\frac{\hat \lambda }{2}\int_{0}^{t^{*}}dt\left(C(t,t)-1 \right)+
\frac 1 2\tr\ln M~,
\end{aligned}
\end{equation}
where the operator $M$ reads
\begin{equation}
\label{M}
M= 
\left(\begin{array}{cc} 
R(t,t')& C(t,t')\\
D(t,t')& R(t',t)
\end{array} \right). \qquad 
\end{equation}
The trace is over times and components. Note that we have {\em two} Lagrange multipliers $\lambda(t)$ and $\hat \lambda(t)$. The corresponding saddle-point equations are shown below, see also Ref.~\cite{BiroliKurchan01}.

One has to find periodic dynamic solutions of period $T_s/(2T_q)$  which for $T_s<T_d$
(a) break causality, (b) have non-zero action, and (c) satisfy time-translational invariance.  
The solution for $T_q\rightarrow 0$ was worked out in Ref.~\cite{BiroliKurchan01}. It leads to the result that the trace over periodic trajectories is equal to the number of states with infinite lifetime, which was previously obtained through the TAP equations~\cite{Rieger1992,KurchanParisiVirasoro,Crisanti1995}.
The analysis of Ref.~\cite{BiroliKurchan01} confirms what we anticipated above.  In particular, it demonstrates that the zero-temperature entropy of the quantum problem is finite (and equal to the complexity) and that the quantum dynamics at $T_q=0^+$ is critical. In order to obtain information on how criticality is cut off and the values of the critical exponents, one has to go beyond this analysis 
and study small but finite $T_q$. A complete ansatz for this regime has yet to be found. In the following we present two approximations and discuss later their limitations. 

\subsection{Equations}

The conditions for stationarity of the action are equivalent to four
 equations for the two-time functions (see Ref.~\cite{BiroliKurchan01}, in Appendix~\ref{app:susy} we review the superspace notation
 that helps simplify these calculations). With $k=p(p-1)/2$,
\begin{equation}
\underline{\partial_t C(t,t')}=-\lambda(t) C(t,t')+ 2T R(t',t)
+\frac{p}{2}\int_{0}^{t^{*}}\di t'' C^{p-1}(t,t'') R(t',t'') 
+\kp\int_{0}^{t^{*}}
R(t,t'') C^{p-2}(t,t'') C(t'',t') \di t''~,
\label{C-eq}
\end{equation}

\begin{equation}
\begin{split}
\underline{\partial_t R(t,t')}=&-\lambda(t)  {R(t,t')}+ 
\underline{2T D(t,t') }+ \frac{p}{2}\int_{0}^{{t^{*}}}\di t''C^{p-1} (t,t'') D(t'',t')\\
&+ \kp \int_{0}^{t^{*}}\di t'' C^{p-2} (t,t'') R(t,t'')R(t'',t')+\delta (t-t')~,
\end{split}
\label{R-eq1}
\end{equation}
\begin{equation}
\begin{split}
\underline{\partial_t R(t,t')}=&-\lambda(t) R(t,t')+ 
\kp\int_{0}^{{t^{*}}}\di t''D(t',t'')C^{p-2}(t',t'') C(t,t'')
+ \kp  \int_{0}^{t^{*}}\di t''C^{p-2}(t',t'')R(t,t'')R(t'',t')\\
&+\kp (p-2)\int_{0}^{t^{*}}\di t''C^{p-3}(t',t'')R(t',t'')R(t'',t')C(t,t'')
-{\hat {\lambda }(t') C(t,t')}+\delta (t-t')~,
\end{split}
\label{R-eq2}
\end{equation}
\begin{equation}
\begin{split}
\underline{\partial_t D(t,t')}=&\lambda(t) D(t,t')-
\kp \int_{0}^{t^{*}}\di t''D(t',t'')R(t'',t)C^{p-2}(t,t'')
-\kp\int_0^{t^*}\di t''D(t,t'')C^{p-2}(t,t'')R(t'',t') \\
&-\kp(p-2) \int_{0}^{t^{*}}\di t''R(t,t'')R(t'',t)R(t'',t')C^{p-3}(t,t'')
+{\hat {\lambda }(t) R(t,t')}~.
\end{split}
\label{D-eq}
\end{equation}

The spherical condition fixes the values of $\lambda$ and $ \hat \lambda $, which can be 
obtained by subtracting Eq.~\eqref{R-eq2} from Eq.~\eqref{R-eq1} for $t=t'$
\begin{equation}\label{lambdac}
\hat{\lambda}(t)=\frac{p(p-2)}{2}\left(\int_{0}^{t^*}\di t''
C^{p-1}(t,t'')D(t,t'')
+(p-1) \int_{0}^{t^*}\di t''R(t,t'')
R(t'',t)C^{p-2}(t,t'')\right)-2TD(0)~,
\end{equation}
\begin{equation}
    \lambda(t) = \frac{p^2}{2} \int_0^{t^*}C^{p-1}(t,t'')R(t'',t)\di t'' + T \left[R(0^+)+R(0^-)\right]\ .
\end{equation}

\subsection{Reparametrization invariance}

Most terms in the equation above obey a reparametrization invariance which is essentially the one of the aging regime~\eqref{repa1},
\begin{equation}
\begin{aligned}
C(t,t') &\rightarrow C(h(t),h(t'))~,\\
R(t,t') &\rightarrow \dot h(t') \;R(h(t),h(t')) ~,\\
D(t,t') &\rightarrow \dot h(t)\dot h(t') \; D(h(t),h(t')) ~,
\end{aligned}
\label{repa2}
\end{equation}
and
\begin{equation}
\begin{aligned}
\lambda(t) &\rightarrow \lambda(h(t)) ~,\\
\hat \lambda (t) &\rightarrow \dot h(t) \; \hat \lambda (h(t)) ~,
\end{aligned}
\label{repa3}
\end{equation}
with now the added reparametrization of $D$, $\hat \lambda$, which were identically zero in causal cases, but not here.
This invariance is broken by underlined terms in the equations, namely:
\begin{itemize}
\item All derivative terms,
\item The term $2TD(t,t')$ in~\eqref{R-eq1}.
\end{itemize}

Derivative terms are neglected at low frequencies, as usual.
If we assume  $\int_0^{\beta_q}  \; D(t,t')$ is small 
  and then we may neglect all terms breaking reparametrization invariance {\em at long times} in the equation of motion. 
  By the same token, the term in the action
  \begin{equation}
  \int_{0}^{t^{*}}dt\left.\left(\partial_{t}R(t,t')
+\lambda R(t,t')-TD(t,t')\right)\right|_{t'=t^+}
\end{equation}
 may be neglected.  
  Under these stipulations the partition function of our ``quantum'' model has reparametrization invariance, just like in the SYK case.

\subsection{Timescale separation for large \texorpdfstring{$\beta_q$}{betaq} and the residual symmetry}
We shall not try to solve the equations on $C,R,D$ here, but use alternative techniques to study some particular limits in the next  sections. In the following we just discuss what form we expect 
for the solution. 

These equations have been solved by fixing the trajectories at a given value of the potential $V$~\cite{BiroliKurchan01},
and the results concerning the number of metastable states, previously obtained through the TAP equations~\cite{Crisanti1995},  were rederived via a purely dynamic approach. 
Here we are interested  in the {\em total} number of states of given lifetime $\beta_q$, a somewhat different and harder calculation.

We may expect a solution of the form
\begin{equation}
\begin{aligned}
C(t,t') &= C_f(t-t') +{\cal{C}} \left(\frac{t-t'}{\beta_q}\right)~,\\
R(t,t') &= -T_s C_f'(t-t') + \frac{1}{\beta_q} {\cal{R}}  \left(\frac{t-t'}{\beta_q}\right)~,\\
D(t,t') &= \frac{1}{\beta_q^2} {\cal{D}} \left(\frac{t-t'}{\beta_q}\right)~,\\
\lambda(t) &= \lambda~,\\
\hat\lambda(t) &= \frac{1}{\beta_q} \hat \lambda_0~,
\end{aligned}
\end{equation}
 where  $C_f(t)$ is the ultraviolet part, and gives the fast relaxation channel within a metastable state. 
The solution in Ref.~\cite{BiroliKurchan01} is of this form with ${\cal{C}},{\cal{R}},{\cal{D}}$  constants.
 
 \subsection{The residual symmetry}
 
The residual reparametrization symmetries include time-translations, and possibly some residual supersymmetry. However we have not identified any $SL(2)$ subalgebra as there is in the SYK model.
Similarly, the finite $\beta_q$ solution is obtained through stretching, rather than a nonlinear function, as in going from Eq.~\eqref{eq:SYK-G0} to Eq.~\eqref{eq:SYK-Gbeta}.

\section{The \texorpdfstring{$p=2$}{p2} case}
\label{sec:p2}

In the following we consider in detail the $p=2$ case. This is a less interesting case since the $p=2$ model is essentially quadratic and falls outside of the class of glass models ($p>2$) which embody the properties focused on in the previous section. The exercise is however instructive to see how the mapping works and to spell out some simple results that will be useful for the analysis of the $p=3$ case analyzed later. 

	For $p=2$, both the original (classical) and the modified (quantum) potentials are quadratic forms in the coordinates.
	The classical system undergoes linear stochastic dynamics, and there is no truly glassy phase with many metastable states, although there is a phase transition to a low-temperature regime where equilibration time is infinite.
	The corresponding quantum model is a set of harmonic oscillators, aside from the coupling arising from the spherical constraint.
	However the physics of the $p=2$ model is not completely trivial: it has a transition at $T_q=0$ where it becomes gapless and the correlation time
	diverges as a power law as $T_q^{-2}$. 
	It is hence worth presenting it as an introduction to the more complex $p>2$ case. 	
	
	\subsection{The model}
	For $p=2$ (linear dynamics), the effective potential is expressed as the quadratic form
	\begin{equation}
	V_{\textrm{eff}} = \frac{1}{2} (\mathbf{q}, A \mathbf{q}) -\frac{1}{4} N T_s \lambda~, \qquad A = \frac{1}{4}(\mathbf{J}-\lambda\mathbb{I})^2\ .
	\end{equation}
	The system is a collection of harmonic oscillators, corresponding to the eigenvectors of $A$, independent except for the spherical constraint
	$\sum_i \braket{q_i^2} = N$, which fixes the Lagrange multiplier $\lambda$.
	
	The oscillators have frequencies $\omega_\mu = |\mu-\lambda|/2$, where $\mu$ are the eigenvalues of $\mathbf{J}$.
	Up to subleading corrections, $\mathbf{J}$ is a GOE random matrix, so in the thermodynamic limit 
	the distribution of $\mu$'s is the Wigner semicircle law of radius $R=J$:
	\begin{equation}
	\rho(\mu) = \frac{2}{\pi R^2} \sqrt{R^2 - \mu^2}\; \theta(R-|\mu|)\ .
	\end{equation}
	The density 
	of oscillator frequencies is simply related to this distribution
	$$\rho(\omega)=\int d\mu \rho(\mu) \delta(\omega- |\mu-\lambda|/2).$$
	
	\subsection{Thermodynamics}
	The partition function at temperature $T_q=\beta_q^{-1}$ is
	\begin{equation}
	Z = \prod_{\mu}\left(\frac{\e^{-\beta_q T_s \omega_\mu/2}}{1-\e^{-\beta_q T_s \omega_\mu}}\right)
	\e^{N\beta_q T_s \lambda/4}~.
	\end{equation}
	The Lagrange multiplier $\lambda$ is fixed by the spherical constraint
	\begin{equation}
	\sum_{\mu}\braket{q_\mu^2} = \sum_\mu \frac{T_s}{2\omega_\mu} +
	\sum_{\mu}\frac{T_s}{\omega_\mu} \frac{\e^{-\beta_q T_s\omega_\mu}}{1-\e^{-\beta_q T_s\omega_\mu}}
	\overset{!}{=}N\ .
	\end{equation}
	We assume that no oscillator is macroscopically occupied, \emph{i.e.} that the $\braket{q_\mu^2}$'s do not
	diverge with $N$. Then in the thermodynamic limit
	the constraint can be expressed in terms of the integral
	\begin{equation}
	\label{eq:Fint1}
	\frac{1}{T_s} =
	\int \di \mu \frac{\rho(\mu)}{|\lambda-\mu|} \coth\left(\beta_q T_s \frac{|\lambda-\mu|}{4}\right)\ 
	\equiv F(\lambda)~.
	\end{equation}
	
	\begin{itemize}
	\item \textbf{Zero-temperature case:}
	For $T_q=0$ the previous equation simplifies to
	\begin{equation}\label{eq:Fint2}
	\frac{1}{T_s} = 
	\int \di \mu \frac{\rho(\mu)}{|\lambda-\mu|}~. 
	\end{equation}
	A solution, $\lambda>R$, is found for $T_s>T_c={R}/{2}$. Instead, for $T_s<T_c$ one has to take into account the appearance of a zero mode in $A$, which is macroscopically occupied.
	This is the same mechanism that leads to Bose-Einstein condensation, although the constraint is different.
	To treat this, we consider that $\lambda$ is a distance $1/N$ to the largest eigenvalue of the matrix $\mathbf J$, 
	and re-write the spherical constraint as
	\begin{equation}\label{eq:spher-bec}
	N \overset{!}{=} N q +\sum_{\mu\neq 0} \frac{T_s}{2\omega_\mu}
	\xrightarrow{N\to\infty} N [q + T_s F(\lambda)] 
	\end{equation}
	and obtain $q=1 - \frac{T_s}{T_c}$, which corresponds to a condensation into the lowest energy mode. 
	Note that with the usual conventions, $T_c = R/2 = J/2$, so this condensation
	is a quantum phase transition that takes place at strong coupling, $J > 2 T_s$.
	For $T_s<T_c$, the density of oscillator is simply a shifted semi-circle with support $[0,2R]$. The spectral density $\rho_q(\omega)=\rho(\omega)/2\omega$ therefore diverges as $1/\sqrt{\omega}$ as small $\omega$. It is also possible to show that the zero-temperature entropy is equal to zero. 
	
	\item \textbf{Finite temperature case:} For $T_q>0$ Eq.~\eqref{eq:Fint1} always has a solution $\lambda>R$ for any $T_s$ since now the integral has a divergence for $\lambda \rightarrow R$. The analysis of Eq.~\eqref{eq:Fint1}
	for $T_q\rightarrow 0$ is slightly involved and can be found in Appendix~\ref{app:p2}. Calling $z=\lambda-R$ one finds that 
	$z$ tends to a finite positive value for $T_s>T_c$, it scales as $z\sim T_q$ for $T_s=T_c$ and  as $z\sim T_q^2$ for $T_s<T_c$. The scaling of $z$ with temperature is important to establish the behavior of the specific heat. In fact, for $T_s>T_c$ 
	a finite $z$ implies a gap in the spectrum for $T_q=0$ and hence an exponentially small specific heat, whereas  
	$\rho(\omega)$ has a gap for $T_s\le T_c$ which scales to zero faster (for $T_s<T_c$) or at the same speed (for $T_s=T_c$) than $T_q$. Given that all oscillators up to frequencies of the order of $T_q$ are excited, their density is $T_q^{3/2}$, 
	and each one gives a contribution of the order $T_q$. Thus one finds an average thermal energy that scales as $T_q^{5/2}$ and a specific heat 
	that scales as $T_q^{3/2}$. A precise derivation is presented in Appendix~\ref{app:p2}. 
\end{itemize}

\subsection{Dynamics}
To study the real-time dynamics we consider the correlator
\begin{equation}
	C(t) = \frac{1}{2 N} \sum_i\Braket{\left\{q_i(t), q_i\right\}}
	= \frac{1}{N} \sum_{\mu}\braket{q_\mu^2} \cos(\omega_\mu t)~,
\end{equation}
where the expectation values are computed in the thermal state of the harmonic oscillators and $\{,\}$ is the anticommutator.

\begin{figure}
	\centering
	\includegraphics[width=.48\textwidth]{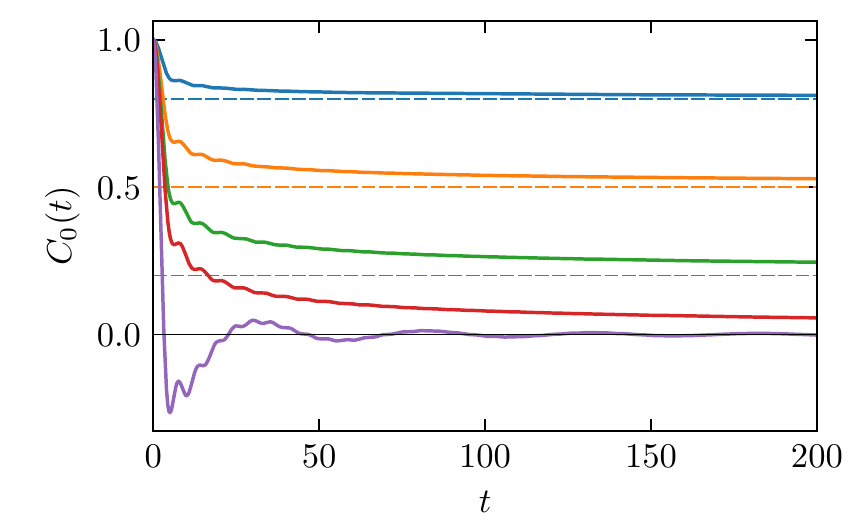}
	\includegraphics[width=.48\textwidth]{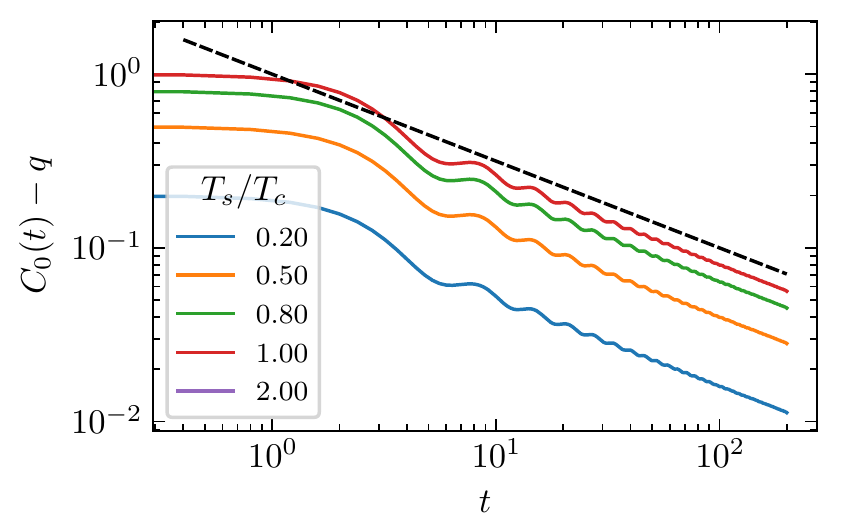}
	\caption[Correlation functions at $T_q=0$ ($p=2$)]{Correlation functions at $T_q=0$ in the $p=2$ model~\eqref{eq:p2-correlator-t0}, for some values of $T_s$ (legend on the right).
		Left: the correlation functions approach the plateau (dashed lines) for $T_s<T_c$,
		and zero for $T_s\geq T_c$. Right: scaling above the plateau for $T_s\leq T_c$,
		showing the $t^{-1/2}$ power law (black dashed line).}\label{fig:syk-p2-c0}
	\end{figure}
	
Taking into account the macroscopically occupied zero-mode, in the thermodynamic limit the correlator has the integral representation
\begin{equation}\label{eq:p2-correlator-t0}
	C(t) = \frac{1}{N}\braket{q_0^2} + T_s \int \di \mu \rho(\mu) \braket{q_\mu^2}\cos(\omega_\mu t) 
	= q + T_s \int \di \mu \frac{\rho(\mu)}{\lambda-\mu} \cos\left(\frac{\lambda-\mu}{2} t\right)~.
\end{equation}

Let us first focus on $T_q=0$. Above the classical transition ({$T_s>T_c$}) 
there is no condensation, $q=0$, and the gap in $\rho(\omega)$ leads to the behavior 
\begin{equation}\label{eq:p2-corr-above}
	C(t) \approx \frac{\cos(\omega_\textrm{min} t)}{t^{3/2}},\quad 
    \omega_{\textrm{min}} = \frac{z}{2} = \frac{1}{2T_s} \left(T_s-T_c\right)^2 \ .
\end{equation}
Below the transition ({$T_s<T_c$}) 
the square root singularity of $\rho(\omega)$ in zero leads to the behavior
\begin{equation}\label{eq:p2T0asympt}
    C(t) = q + \frac{b}{t^{\frac{1}{2}}}\ .
\end{equation}
The power law decay is the same at and below the transition $T_s\leq T_c$. See Fig.~\ref{fig:syk-p2-c0}.

The properties of $\rho(\omega)$ also fix the behavior of the imaginary time correlator $C_q^I$ defined in Eq.~\eqref{cqi}. It displays an exponential relaxation for {$T_s>T_c$} and a power law approach to $q$ analogous to~\eqref{eq:p2T0asympt} for {$T_s<T_c$}. As expected, this is exactly the same behavior of the stochastic correlation function. 
Thus the criticality, and its absence, in the quantum problem can be directly traced back to the dynamical critical behavior (and its absence) for stochastic dynamics~\cite{CugliandoloDean,Ciuchi1988}.

The results for $T_q\rightarrow0$ can also be understood from the properties of $\rho(\omega)$ at non-zero temperatures. For $T_s>T_c$, the finite gap in the spectrum leads to the same behavior obtained at $T_q=0$ for real and imaginary time correlators. For $T_s<T_c$ there is gap but it scales as $T_q^2$. As a consequence, in real time, the correlator shows an intermediate regime $1\ll t \ll \beta_q^2$ in which it
approaches the constant value $q$, with a $t^{-\frac{1}{2}}$ power-law decay, but then at $t\propto \beta_q^2$ the correlator decays from the	 plateau to zero. In the imaginary time this second regime is instead invisible since times are bounded by $1/T_q$ and thus ones finds a power law approach to the plateau and then a mirror image for $\tau>\beta_q/2$ as a consequence of periodicity.
A more detailed derivation of all these results is presented in Appendix~\ref{app:p2}. 

\subsection{Summary}
The results found for the $p=2$ (summarized in Table~\ref{tab:p2}) case show some of the properties of the SYK model but not all. The specific heat displays a non-trivial scaling for $T_q\rightarrow 0$ but the zero-temperature entropy is zero and the relaxation time scale diverges at zero temperature as $\beta_q^2$ and not $\beta_q$. However, we see at play some of the ingredients that will emerge as important in the analysis of the $p>2$ case. The criticality (power-law behavior) of the zero-temperature quantum dynamics is directly related to the criticality of the corresponding stochastic dynamics for $T_s<T_c$. Moreover, the effect of a finite small temperature is to select  stochastic dynamics trajectories, i.e. in imaginary time trajectories for the quantum problem, which explore the part of configuration space dominated by metastable states with a finite lifetime. The lifetime is directly related to the gap in the spectrum of harmonic oscillators, which scales as $T_q^2$ for $p=2$. The vanishing zero-temperature quantum entropy is directly related to the 
number of long-lived metastable states, since this is not exponential in $N$ for the $p=2$ model, the zero-temperature entropy vanishes for $T_q\rightarrow0$. In order to obtain a different result one has to consider classical models with a much rougher energy landscape. This is what we do in the following focusing on $p>2$. 

\begin{table}
    \caption{Summary of results for $p=2$.}\label{tab:p2}
	\renewcommand{\arraystretch}{1.6}
	{\centering
	\begin{tabular}{c|c|c|c}
	    \hline \hline
		& $T_s<T_c$ & $T_s=T_c$ & $T_s>T_c$ \\ 
		\hline 
		$q$ & $1-T_s/T_c$ & 0 & 0 \\ 
		$z$ & $T_q^2$ & $T_q$ & $(T_s-T_c)^2/T_s$ \\ 
		specific heat & $ T_q^{3/2}$ & $ T_q^{3/2}$ & $\e^{-\beta_q T_s z/2}$ \\
		dynamics & Plateau $q+\frac{b}{t^\frac{1}{2}}$ for
		$1\ll t\ll \beta_q^2$ & $\frac{b}{t^\frac{1}{2}}$ for
		$1\ll t\ll \beta_q$ &
	   	$\e^{-\imi zt/2}/t^{3/2}$ \\
	    \hline \hline
    \end{tabular} }
\end{table}

\section{The \texorpdfstring{$p \ge 3$}{p3} case}
\label{sec:p3}

\subsection{The general picture}  

Armed with what we have learned from the solution of the $p=2$ model and using the general results from the mapping between stochastic and quantum dynamics, we are now in a position to study the low $T_q$ behavior of the quantum model $H$, the counterpart of the Fokker--Planck operator of the $p=3$ spherical $p$-spin model.

Henceforth, we consider $T_s<T_d$ so that metastable states are well formed.  
In computing the partition function, we are summing over all metastable states of lifetime $1/T_q$. Now, from what we know concerning the 
metastable state distribution Fig.~\ref{fig:complexity}, the vast majority of stable (in the limit $N \rightarrow \infty$) states fall just below the threshold level. (For this discussion terms such as  ``high,'' ``low,'' ``above,'' and ``below'' refer to the original energy $V(q)$ in the classical model and not the effective potential $V_{eff}$ of the quantum model).
For small but non-zero $T_q$ just above the threshold
there are many more states with finite lifetime diverging as $1/T_q$, with the higher states having shorter lifetimes. There is thus a tradeoff,
and the natural result is that the temperature $T_q$ selects the highest---and hence more numerous---metastable states with lifetime $1/T_q$. As $T_q \rightarrow 0$ the best one can do is work right at the threshold. 
As $T_q \rightarrow 0$ the threshold is asymptotically approached. 
Only when $1/T_q$ diverges with $N$ first as a power law and then as an exponential, metastable states with less stability are excluded. Therefore, we note the important fact that at $T_q=0^+$ \emph{the entropy of the quantum model is finite} and precisely equal to the complexity of the most numerous metastable states with an infinite lifetime, namely the threshold states (recall that the energy is related to the inverse of the relaxation time, hence it is zero for those states). 

The second important fact of note is that threshold states are marginal and thus their Hessian is gapless. As a consequence, the stochastic correlations, as well as the quantum imaginary and real time correlations, have a power law behavior in time approaching a finite overlap $q$. At finite but small $T_q$ the stochastic trajectories correspond to metastable states with lifetime $1/T_q$, and criticality is expected to be cut off. Therefore \emph{$T_q=0$ is a quantum critical point} at which we expect critical thermodynamic and dynamical behavior. In order to obtain the critical exponents of the vanishing specific heat and the divergent relaxation time a detailed analysis is needed. We develop the framework to perform it below, and present a first step toward a complete solution.

\subsection{Two simple approximations}

The saddle-point equations simplify when the $T_s\rightarrow 0$ limit is taken simultaneously with the $T_q\rightarrow 0$ one, 
as shown in~\cite{BiroliKurchan01}. Our first step is therefore to analyze the limit $T_s\to 0$, $T_q\to 0$ at fixed $t^* = T_s/(2T_q)$, and analyse the scaling with $t^*\to \infty$.
	This provides a first approximation, but is different from the $T_q\to0$ limit at fixed $T_s$. From the point of view of the classical model, it allows one to study the long-time dynamics at zero classical temperature.
	
	At $T_s\to 0$, the trace over periodic trajectories at classical energy $\mathcal{E}$, or equivalently the entropy density of states of energy $\mathcal{E}$ stable up to $t^*$ is given by~\cite{BiroliKurchan01}
	\begin{multline}\label{eq:p3-entropy-t0}
	s(\mathcal{E},t^*) = \frac{1}{2}\left(1+\ln\frac{p}{2}\right) - \mathcal{E}^2 + \re\left[\frac{1}{2}\left(\frac{\mathcal{E}\mp\sqrt{\mathcal{E}^2-\ethr^2}}{\ethr}\right)^2
	+ \log\left(-\mathcal{E}\mp\sqrt{\mathcal{E}^2-\ethr^2}\right)\right]+\\
	-\int \di \omega \rho_p(\omega+p\mathcal{E}) \ln\left[1-\e^{-t^*|\omega|}\right]
	+ t^* \int_{-\infty}^{0} \di\omega \omega \rho_p(\omega+p\mathcal{E})~.
	\end{multline}
	The integrals involve $\rho_p$, the semicircle density of radius $R=\sqrt{2p(p-1)}$, centred at $-p\mathcal{E}>0$.
	
\begin{figure}
		\centering
		\includegraphics{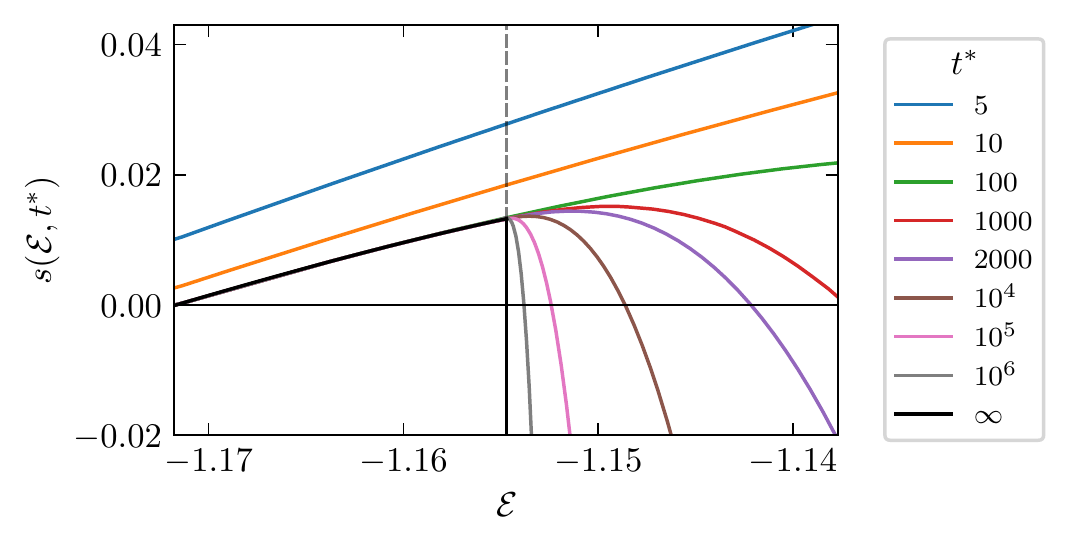}
		\caption[Time-dependent configurational entropy in the classical 3-spin model]{Time dependent, energy-resolved configurational entropy at $T_s\to0$ for $p=3$, reproducing Fig.~4 of Ref.~\cite{BiroliKurchan01}.}\label{fig:syk-BK4}
	\end{figure}

	The first line of~\eqref{eq:p3-entropy-t0} does not depend on $t^*$ and counts the number of saddles (stationary points in the energy landscape) at energy density $\mathcal{E}$.
	The second line is a sum of harmonic contributions, and the density of states $\rho_p$
	coincides with the spectrum of the Hessian computed at saddles of energy density $\mathcal{E}$~\cite{KurchanParisiVirasoro,Cavagna1998}. It is interpreted as a harmonic expansion around the saddles.\footnote{The expansion becomes exact at $T_s\to 0$~\cite{BiroliKurchan01,Kurchan09-lectures}. This is the idea behind the harmonic approximation presented in the next section.}
	As we show below, if $\rho_p$ has positive support, the contribution from the second line is vanishingly small at large $t^*$;
	otherwise, it gives an increasingly negative contribution, a penalty for expanding around unstable saddles. 
	The energy at which the edge of the semicircle touches zero is the threshold $\ethr = -\sqrt{2(p-1)/p}$.
	In Fig.~\ref{fig:syk-BK4} we show the configurational entropy~\eqref{eq:p3-entropy-t0} as a function of~$\mathcal{E}$, for increasing values of $t^*$.

	To recover the partition function of the quantum model, we are interested in the \emph{total} number
	of metastable states at $t^*$, regardless of energy.
	In terms of entropy, this is controlled for each $t^*$ by the maximum over $\mathcal{E}$ of~\eqref{eq:p3-entropy-t0}.
	
	For increasing $\mathcal{E}$ at fixed $t^*$, there is a competition between the two terms:
	the total number of saddles increases, while $\rho_p$ shifts towards negative values, making the contribution from the integrals more negative.
	In the $t^*\to \infty$ limit the number of stable states is recovered (black line in Fig.~\ref{fig:syk-BK4}), in agreement with the TAP calculation~\cite{Crisanti1995}, and
	the maximum is at the threshold $\ethr$, with configurational entropy $s_0=s(\ethr, \infty)$.
	For finite $t^*$, there is a unique maximum $\mathcal{E}_M(t^*)$, which approaches the threshold from above.
	
	We are interested in the scaling of $\mathcal{E}_M(t^*)-\ethr$ and of $s_M(t^*)-s_0$ with $t^*$.
	To determine these scalings, we consider~\eqref{eq:p3-entropy-t0} in the double scaling limit $t^*\to \infty$, $\mathcal{E}\to\ethr$ with $\mathcal{E}-\ethr = A t^{-\alpha}$ for some fixed $\alpha$.
	We then determine the exponent $\alpha$ by comparing the competing contributions in~\eqref{eq:p3-entropy-t0}.
	The calculation is performed in Appendix~\ref{app:p3}. We find the exponent $\alpha=2/3$ independent of $p$, and
	\begin{equation}\label{eq:t0-sscaling}
	s_{M}(t^*) = s\left(\mathcal{E}_M(t^*),t^*\right) = s_0 + c_M t^{*-\frac{2}{3}} + \mathcal{O}\left(t^{*-\frac{4}{3}}\right)
	\end{equation}
	with a $p$-dependent constant $c_M>0$, see Eq.~\eqref{eq:t0-smax}.
	
	Using the correspondence between the number of metastable states in the classical model and the partition function of the quantum model, we derive from~\eqref{eq:t0-sscaling} the free energy of the latter at $T_q\to 0$
	\begin{equation}\label{eq:t0-free-energy}
	-\beta_q f = \frac{1}{N}\ln\mathcal{N}\left(t^* = \frac{T_s}{2T_q}\right) = s_0 + 2^{-2/3}c_M \left(\beta_q T_s\right)^{-2/3}\ .
	\end{equation}
	This shows that the model has finite entropy $s_0$ at zero temperature. Like in the SYK model, this is not due to degeneracy
	(the ground state is unique for any finite $N$), but to the ``accumulation'' of an exponential number of stable states at the threshold. From~\eqref{eq:t0-free-energy} we also derive the scaling of the energy density $\propto T_q^{5/3}$
	and specific heat $\propto \left({T_q}/{T_s}\right)^{2/3}$.
	Thus, we have reobtained a similar critical behavior of the $p=2$ case but with a finite entropy at zero temperature. The states contributing to the entropy dominate the low-temperature specific heat, changing the exponent from that of the $p=2$ case.
	Clearly the specific heat exponent differs somewhat also from that of the SYK case.
	
	In order to go beyond this first approach, we consider the low-$T_q$ scaling at fixed small $T_s$, using a harmonic expansion for the low $T_{s}$ dynamics, which consists in expanding the potential around each stationary point and approximating the degrees of freedom as harmonic oscillators, with frequencies given by the spectrum of the Hessian.
	This expansion includes unstable directions, whose effect is taken into account in the resulting spectrum.
	The expansion is presented and discussed in Chapter 3 of Ref.~\cite{Kurchan09-lectures}.
	As $T_s\to 0$, the expansion becomes exact and the result~\eqref{eq:p3-entropy-t0} is recovered, while for small $T_s>0$ it provides an approximation only.
	The computation is presented in Appendix~\ref{app:p3}. The final result for the entropy, free energy and specific heat displays the same scaling with $T_q$ found above.  Within the harmonic approximation one can also obtain the quantum correlation
	functions (for $T_s=0$ these are trivial since $q=1$). As discussed in Appendix~\ref{app:p3}, the result is analogous to the $p=2$ case but with a different spectral density $\rho (\omega)$. 
	\begin{equation}\label{eq:p3-corr}
	C(t) = \frac{1}{N}\sum_i\braket{\{q_i(t),q_i\}} = \int \di\omega \frac{\rho(\omega)}{2\omega} \coth\left(\frac{\beta_q T_s}{2} \omega\right) \cos(\omega t)~,
	\end{equation}
	with
	\begin{equation}\label{eq:deformed-DoS}
	\rho(\omega) = \int \di\rho_p(\mu) \delta\left(\omega- \sqrt{\frac{1}{2}\hat{\lambda}T_s+\frac{1}{4}(\lambda-\mu)^2}\right)\ .
	\end{equation}
	At $T_q=0$, $\lambda=R$ and $\hat{\lambda}=0$, and the critical behavior is the same as for $p=2$. Note that the critical temperature $T_c$ is rescaled and $p$-dependent; since we are working at small $T_s$, we are deep in the condensed phase, and $q=1-T_s/T_c$ is close to one.
	
	\begin{figure}
		\centering
		\includegraphics[width=.4\textwidth]{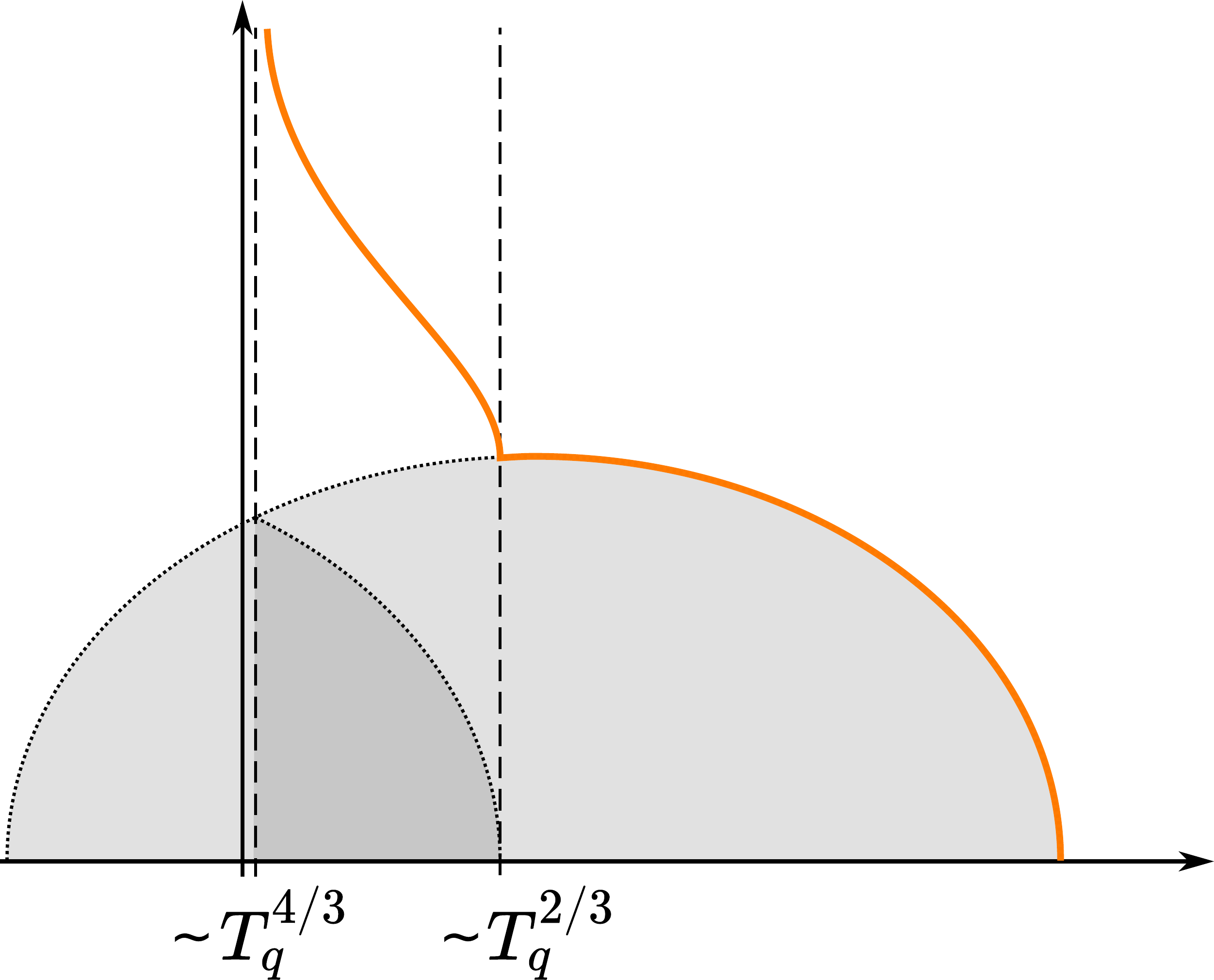}
		\caption[Deformed density of states in the harmonic approximation]{Sketch of the density of states~\eqref{eq:deformed-DoS} for $p>2$, $T_s>0$ within the harmonic approximation (distribution of $\omega$, orange line) compared to the $T_s=0$ semicircle (distribution of $(\lambda-\mu)/2$).}
		\label{fig:deformed-sc}
	\end{figure}
	
For $T_q>0$, given the semicircle-distributed spectrum for $\mu$, a change of variables leads to the deformation sketched in Fig.~\ref{fig:deformed-sc}.
There are two relevant scales, both vanishing in the $T_q\to 0$ limit: $z=(R-\lambda)/2\approx T_q^{2/3}$ and $\omega_{\textrm{min}}=\sqrt{\hat{\lambda} T_s/2} \approx T_q^{4/3}\ll z$.
For $\omega\geq z$, there is a one-to-one correspondence between $\omega$ and $\mu$, and the distribution $\rho(\omega)$ is very close to the semicircle centred in $\lambda$ for $\omega\geq z$.
For lower $\omega$, each value of $\omega$ is obtained from two different $\mu$'s with the edge of the semicircle ``folded back'' to positive values, giving a square-root kink at $\omega=z$. Finally, $\omega_{\textrm{min}}$ acts as a cut-off.
	
As for $p=2$, the behavior of $\rho(\omega)$ at small but finite $T_q$ allows one to obtain the long time behavior of the correlation functions. The real time quantum correlation function is the same as the $T_q=0$ up to time scales $t \lesssim z^{-1}\approx T_q^{-2/3}$ where the plateau is approached, for which the system cannot resolve the difference between the two densities of states.
The departure from the plateau takes place on the timescale $\propto T_q^{-4/3}$, set by the gap $\omega_{\textrm{min}}$, at which the correlation function decays exponentially. In the intermediate regime between those timescales $C$ remains close to the plateau up to terms vanishing as power laws of $T_q$.
As for the imaginary time quantum correlation function, one observes only the first power law relaxation toward the plateau which is then folded back due to periodicity. The second regime is invisible since it corresponds to frequencies much smaller than $1/\beta_q$.

\vspace{\baselineskip}

\begin{table}
    \caption{Summary of results for $p\geq3$ ($T_s\to0$).}\label{tab:p3}
	\renewcommand{\arraystretch}{1.6}
	{\centering
	\begin{tabular}{c|c}
	    \hline \hline
	    entropy & $s_0=\Sigma(\ethr)$\\
       	\hline 
		$z$ & $T_q^{2/3}$ \\ 
		\hline 
		specific heat & $ (T_q/T_s)^{2/3}$ \\
		\hline
		gap & $\omega_{\textrm{min}} \propto T_q^{4/3}$\\
		\hline
		$q$ & $1-T_s/T_c$ \\ 
		\hline
		dynamics & Plateau $q+\frac{b}{t^\frac{1}{2}}$ for
		$1\ll t\ll \beta_q^\frac{4}{3}$ \\
	    \hline \hline
    \end{tabular} }
\end{table}

In conclusion, within the approximations presented in this section, we obtain many of the desired features of the SYK model,
see Table~\ref{tab:p3}, in particular a quantum critical point at $T_q=0^+$ with finite entropy.  It remains to be seen whether 
the critical behavior found within the harmonic approximation is representative of the result for small but finite $T_s$. The main concern is that the periodic trajectories are extremely simple within these approximations and do not explore at all 
the rough landscape but remain very close to a given critical point.

\section{Generalizations}
\label{sec:generalisations}
We now discuss three questions that have arisen naturally in the context of the SYK model through the lens of classical glassy dynamics mapping. 

\vspace{.2cm} 

$\bullet$ {\em Transition to normal quantum liquid.}

\vspace{.2cm} 

Banerjee and Altman~\cite{BanerjeeAltman} showed that perturbing the SYK Hamiltonian with a quadratic term, the zero-temperature
entropy is a decreasing function of the strength, and reaches zero at a critical value, at which the system becomes gapped. Here the same situation arises naturally. Consider the original diffusive model, perturbed with a ``magnetic field'' term $b \sum_i q_i$. We know~\cite{Crisanti1995} that the number of stable metastable states is a decreasing function of $b$, and reaches zero at a critical field $b_c$, which depends on $T_s$. Above this critical field strength, the system
is no longer glassy, and there are no slow relaxations. This implies for the ``quantum'' associated model that the zero-temperature $T_q$ entropy is a decreasing function of
$b$, and that above $b_c$ the system becomes gapped: the gap being the inverse of the slowest relaxation time.

More generally, one can consider ``mixed'' models, with multiple random couplings with different values of $p$.
This modifies quantitatively (and to a certain extent qualitatively) the dynamics of the glassy model: it is still glassy but for example, some scaling exponents change \cite{CrisantiLeuzzi2007}. This induces 
quantitative (and possibly also qualitative) changes in the ``quantum'' model.
This is unlike the SYK model, where only the term with the smallest $q$ is relevant and dominates at long times.

\vspace{.2cm} 

$\bullet$ {\em Nearby replica symmetry breaking transition.  }

\vspace{.2cm}

 Consider now adding a term to $H$ proportional to the potential:  
\begin{equation}\label{eq:FP-to-H1}
	H_\mu = \sum_i\left[-\frac{T_s^2}{2} \frac{\partial^2}{\partial q_i^2}
	+\frac{1}{8}\left(\frac{\partial V}{\partial q_i}\right)^2 -\frac{T_s}{4}\frac{\partial^2 V}{\partial q_i^2}\right] + \mu V(q)~.
	\end{equation}
	$H$ has still the form of a Schrodinger operator with $T_s$ playing the role of $\hbar$, but now the potential
	is modified as
	\begin{equation}
	V_{\textrm{eff}} = \frac{1}{8}\left(\frac{\partial V}{\partial q_i}\right)^2 -\frac{T_s}{4}\frac{\partial^2 V}{\partial q_i^2}+ \mu V(q)~.
	\label{modif}
	\end{equation}
	From the form of~\eqref{modif} we can already see that the degeneracy between saddles is broken.  Indeed, if we consider the eigenstates of $H$ that are quasi-degenerate
	(their value scales with $N$ in a manner slower than $N$), then the term $\mu V(q)$ is the only relevant term and lifts the degeneracy. In fact, the partition function is then the one of a classical $p$-spin model with inverse temperature $\beta_q \mu$. 
	The system then has a transition temperature at $T_q^{\textrm{crit}} = \mu {T_k}$, where $T_k$ is the thermodynamic transition temperature of a classical $p$-spin spin glass~\eqref{spherical},
	 at which the Gibbs measure freezes in the ground state.
Note that~\eqref{eq:FP-to-H1} with $\mu \neq 0$ no longer corresponds to a diffusive problem, but rather to a diffusive problem with branching proportional to $V$~\cite{Kurchan09-lectures}.

\vspace{.2cm} 

$\bullet$ {\em Models without disorder.}

\vspace{.2cm} 

The question of substituting a disordered model by one with similar phenomenology
but with deterministic Hamiltonian arose in the '90s in the portion of the glass community working with
mean-field models. Several Hamiltonians were proposed, and techniques were developed 
to obtain their disordered  counterparts having the same dynamics.
By considering the evolution operator of any of the models developed then, we obtain
a "quantum" version of strange liquid without disorder, in the spirit of Ref.~\cite{Witten2016tensor}.
 Most of the models we shall describe have  $\pm 1$ spins: we may make them continuous by
 using a "soft spin" version with the addition of a term $\propto \sum_i (s_i^2-1)^2$ to the Hamiltonian,
 or simply directly use Glauber dynamics for Ising spins---the evolution operator of which may also
 be also represented by a Hermitian quantum-like operator.
 Some examples are
 
 \begin{enumerate}
\item The Bernasconi model, originating in information theory~\cite{Bouchaud1994,Marinari1994}:
\begin{equation}
    E= \sum_{k=1}^N \left(\sum_{i=1}^{N-k} s_i s_{i+k} \right)^2~.
\end{equation}
\item The ``Sine'' model~\cite{Marinari1994b}:
\begin{equation}
    E= \sum_{k=1}^N \left[\sum_{i=1}^{N} \frac{1}{\sqrt{N}} \sin \left(\frac{2 \pi ik }{N} s_i\right)-s_k \right]^2~. 
\end{equation}
\item The Amit--Roginsky model~\cite{Franz1995}, a $3$-spin model as~\eqref{spherical},
with $J_{ijk}$ a $3-j$ symbol, rather than random.  Interestingly this model may be viewed as a classical cousin of Witten's tensor generalization of the SYK model~\cite{Witten2016tensor}.
\item A matrix model~\cite{Matrix}, with permutation rather than rotational invariance:
\begin{equation}
  E = \frac{1}{N} \sum_{ab} \left(S^a \cdot S^b \right)^p,
\end{equation}
where $S^a$,  $a=1,...,N$ are  $N$-dimensional vectors with $\pm 1$ entries.
\end{enumerate}
Intriguingly, these models have a landscape with essentially the same density
of minima as their random counterparts, but the few lowest states are exceptional and
related to number theoretic properties of the specific energy functions. We may think of these as the "crystals" of the problem.

\section{Conclusions}
\label{sec:conclusions}

In this work we have embarked on a program to investigate and explore connections between quantum SYK-like models and a broad class of classical glass models.  Specifically, mean-field glassy systems  which have obvious similarities to the SYK class of models already at the level of the Hamiltonian, exhibit deep connections with SYK when viewed from the standpoint of their {\em dynamically critical} behavior.  We have focused on the $p$-spin spherical model but the relationship with glassy dynamics is actually much more general: the evolution operator of any stochastic problem with detailed balance and with a glass transition connected to an exponential number of metastable states presents aspects of SYK-like physics for the reasons we spelled out in this work. The resulting quantum Hamiltonian displays zero-temperature critical behavior with a concomitant finite zero-temperature entropy and time-reparametrization (quasi-)invariance of the dynamical equations for correlations.  These properties have natural classical interpretations.  For example, the dense energy spectrum above the ground state that generates the finite entropy in the SYK model at $T=0$ can be naturally connected to the dense spectrum of relaxation modes at the ``threshold'' of the energy landscape proximal to a dynamical freezing transition where the configurational entropy of the system jumps to a finite value.  More importantly, time-reparametrization, well-known for many years within the context of classical glasses, is there associated with a defining physical feature of dynamics, namely the phenomena of dynamical heterogeneity where particle motion becomes spatially correlated and an associated dynamical length scale diverges at the critical point as marked by the divergence of a particular class of four-point functions. In this regard the behavior of the SYK model as $T \rightarrow 0$ may be viewed as connected to a ``quantum'' type of dynamical heterogeneity with the divergence of a completely analogous four-point susceptibility.   Such connections, interesting in their own right, may have the practical benefit of widening the class of systems that may serve as appropriate duals for models of black holes.

The euclidean time evolution of the SYK model, as far as we can determine, cannot be mapped onto a diffusive problem, but the possibility remains that some heretofore unknown model with the same properties might. In addition, some features of what we call ``strange quantum liquids'' may differ from those of the SYK model and remain to be carefully explored.  One simple example is the power law decay of correlations, whose exponent is a continuous function of the parameters, unlike those found in the SYK model. 
More importantly, the nature of the time evolution of ``out-of-time-ordered'' correlators and the bound on chaos in these systems demand careful scrutiny.  There are tantalizing hints that these systems will, if not saturate the bound, at least have non-trivial quantum effects on scrambling behavior.  For example, consider Eq.~\eqref{eq:Veff}: the classical portion of $V_{\textrm{eff}}$ is zero at saddles of any index and very small along the gradient lines that connect saddles to other saddles.  Such a ``flat bottomed'' high dimensional space provides a platform for classical chaotic motion even as $T \rightarrow 0$ because of the near-zero energetic cost for trajectory spreading~\cite{Kurchan2016,Bilitewski2018,Scaffidi}.  It has been demonstrated that such systems are prime candidates for maximal quantum chaoticity, displaying a temperature dependence of the Lyapunov exponent $\lambda$ which follows $\beta \hbar \lambda \sim T^{-\alpha}$ with $\alpha \sim \frac{1}{2}$.  Since this behavior violates the bound at low $T$, quantum scattering intervenes to cut off the unlimited growth of $\beta \hbar \lambda$ at its maximal value~\cite{Kurchan2016}.  Interestingly, the second (semi-classical) term in Eq.~\eqref{eq:Veff} provides the first clue as to the quantal mechanism for the reduction of the growth in $\lambda$.  This term, proportional to $\hbar$  (\ie~$T_{s}$), cancels the zero-point energy for stable critical points but additively increases the zero-point energy for unstable saddles (the more so the higher the saddle index), thereby selecting trajectories that ``pass'' low-order saddles.  

In conclusion, we have exposed deep and surprising connections between the behavior of classical glasses and quantum models of the SYK variety.  By doing so, we have introduced a new class of quantum models that are interesting in their own right and may provide future inspiration for developments in, and connections between, classical statistical mechanics as well as in hard condensed matter and high energy physics.  Future efforts will be devoted exploring these connections as well as to providing a deeper understanding of the chaotic properties of these new models.

\begin{acknowledgments}
We would like to thank Yevgeny Bar Lev for discussions and collaboration on this topic in the early stages of this work.
This work was supported by the Simons Foundation Grants No. \#454943 (Jorge Kurchan), \#454935 (Giulio Biroli), \#454951 (David R. Reichman).
DF was partially supported by the EPSRC Centre for Doctoral Training in Cross-Disciplinary Approaches to Non-Equilibrium
Systems (CANES, EP/L015854/1) and the European Research Council (ERC) under the European Union's Horizon 2020 research and innovation programme (grant agreement n$^{\circ}$~723955 - GlassUniversality).
\end{acknowledgments}

\appendix
\section{Supersymmetry}
\label{app:susy}

As is well-known~\cite{Witten1982}, the Hamiltonian~\eqref{eq:FP-to-H} may be promoted to supersymmetric quantum mechanics
via the use of fermionic degrees of freedom, their corresponding spaces, and a term
\begin{equation}
H_{\textrm{SUSY}}= H - \frac{\partial^2 V}{\partial
 q_i \partial q_j } a^\dag_i a_j~.
 \label{hsusy}
 \end{equation}
 Clearly, the total fermion number is conserved, and the original problem is the zero-fermion subspace restriction of
 the full SUSY one. 
  Note that, unlike the supersymmetric versions of SYK~\cite{Murugan2017,Fu2017}, here the bottom state is {\em bosonic}.
 
  There are three reasons why looking at the diffusive problem from this perspective is interesting~\cite{Kurchan1992susy}:
 \begin{itemize}
 \item Supersymmetry implies a relationship between the parameters $C$, $R$ and $D$. They are the {\em equilibrium relations}---namely the fluctuation-dissipation relations and time-translational invariance. The glass transition is, in this language, signalled by supersymmetry breaking. 
\item Time reparametrizations are encapsulated in a single "supertime" reparametrization.
 \item More prosaically, it turns out that taking advantage of the superspace notation makes calculations easier
 and more tractable.
 \end{itemize} 
  We may encode the original variables in a superspace variable:
  \begin{equation}
  \phi_i(1)= q_i(t) + \bar \theta a_i + a_i^\dag \theta + p_i \bar \theta \theta~,
  \end{equation} 
 which leads us to
\begin{equation}
\begin{aligned}
Q(1,2)&=\frac 1 N \sum_i \phi_i(t,\theta,\bar\theta) \phi_i(2) =
C(t_1,t_2) + (\bar \theta_2 - \bar \theta_1)
\theta_2  R(t_1,t_2)
+ \bar \theta_1 \theta_1 R(t_2,t_1) + \bar\theta_1\theta_1 \bar \theta_2 \theta_2 D(t_1,t_2) \\ 
&+  \text{odd terms in the $\bar \theta,\theta$}~.
\end{aligned}
\label{Q12}
\end{equation}
Here $\theta_a$, $\bar \theta_a$ are Grassmann variables, and we denote the full set of coordinates 
in a  compact form as
$1= t_1 \theta_1 \overline\theta_1$, $d1=   dt_1 d\theta_1 d\overline\theta_1$, etc.
The odd and even fermion numbers decouple, so we can neglect all odd terms in $\theta,\bar{\theta}$.
The dynamic action takes the simple form
\begin{equation}
    S[Q]  = -\frac{1}{2}\int \di 1 \di 2 \left[ \Dtwo Q(1,2) + \frac{J^2}{2}Q(1,2)^{p}\right]+
    \frac{1}{2}\int \di 1 Z(1) [Q(1,1)-1] - \frac{1}{2} \tr \ln Q~,
\label{eq:susy-action}
\end{equation}
and the 
associated equations of motion
\begin{equation}
 -\Dtwo Q(1,2) + Z(1) Q(1,2) - \frac{J^2 p}{2} \int d1' Q(1,1')^{p-1}Q(1',2) = \delta(1-2)\ .
\label{eq:susy-eom}
\end{equation}
The Lagrange multiplier in superspace encodes for the two bosonic multipliers
\begin{equation}
Z(1)=\lambda (t)+\bar \theta_1  \theta_1 \hat \lambda (t)~,
\label{eq:susy-z}
\end{equation}
and the kinetic term operator is given by the commutator
\begin{equation}
\Dtwo=
\left[\frac{\partial}{\partial \theta},\left(T_s\frac{\partial}{\partial \bar \theta}-\theta \frac{\partial}{\partial t} \right)\right]~.
\end{equation}
Note how close these are, when written in the appropriate notation, to their SYK counterparts~(\ref{eq:SYK-action},\ref{eq:SYK-eom}).
Reparametrization invariance arises from neglecting the first term in~\eqref{eq:susy-eom}.

In general, to the extent that one is 
allowed to neglect the "small" terms in the infrared, 
Eq.~\eqref{eq:susy-eom} is invariant with respect to any change of ``coordinates''
$t_a$, $\theta_a$, and ${\bar \theta}_a$ ($a=1,2,...$) with 
unit super Jacobian~\cite{CugliandoloKurchan1999}.
This is a large symmetry group, including  the time-reparametrization:
\begin{equation}
t_a \rightarrow h(t_a) 
\; , \;\;\;\;\;\;\;\;\;\;
\theta_a \rightarrow {\dot h}(t_a) \theta_a
\; , \;\;\;\;\;\;\;\;\;\;  
{\bar \theta}_a \rightarrow {\bar \theta}_a
\; .
\label{repp}
\end{equation}
which encapsulates all of Eqs.~(\ref{repa2},\ref{repa3}).

\section{Scaling in the \texorpdfstring{$p=2$}{p2} model}
\label{app:p2}
\subsection{Lagrange multiplier}
\subsubsection{\texorpdfstring{$T_q=0$}{tq0}}
For $T_q=0$ the spherical constraint is given by~\eqref{eq:Fint2}. The integral is well known in random matrix theory, representing the resolvent of Wigner's semicircle distribution~\cite{RMT}
\begin{equation}\label{eq:sphcon-t0}
F(\lambda) = \frac{2}{R^2}\left(\lambda -\sqrt{\lambda^2-R^2}\right) .
\end{equation}
Therefore for $T_s>T_c=R/2$, a solution $\lambda=(T_c^2+T_s^2)/T_s>R$ is found, leading to a positive gap.

On the other hand if $T_s<T_c$, Eq.~\eqref{eq:Fint2} has no solution.
As in Bose-Einstein condensation, to satisfy the constraint we must allow for the lowest energy mode to be macroscopically occupied.
To account for this we take the gap to be $\mathcal{O}(1/N)$, corresponding to a condensation $\braket{q_0^2} = T_s/(\lambda-R) \equiv N q$.
The spherical constraint~\eqref{eq:spher-bec} determines $q=1-T_s/T_c$.

\subsubsection{\texorpdfstring{$T_q\to 0$}{t0} scaling}
At any finite temperature $T_q>0$, $F(\lambda)$ in~\eqref{eq:Fint1} is monotonically decreasing and diverges as $\lambda\to R^+$. Therefore a solution is found for any value of $T_s$. There is no condensation and the spectrum is gapped, $z=\lambda-R>0$.
If $T_s\leq T_c$, the gap closes approaching the critical point $T_q\to 0$. Here we determine the scaling of $z$ with $T_q$, which
governs the critical behaviour of other physical quantities. 

If $T_s<T_c$, Eq.~\eqref{eq:Fint1} can be rewritten
\begin{equation}\label{eq:p2-zscaling-integral}
2 \int_0^{2R} \di x\frac{\rho(R-x)}{z+x} \frac{1}{\e^{\beta_q T_s (z+x)/2}-1} =  T_s^{-1}-T_c^{-1}+\mathcal{O}(\sqrt{z})\ .
\end{equation}
The integral in the left hand side must be of order one.
With a change of variables $x'=x/z$, ignoring constant factors and with $c = \beta_q z T_s /2$, it becomes
\begin{equation}\label{eq:p2-zscaling-zint}
\frac{2}{\pi}\sqrt{2z} R^{-\frac{3}{2}} \int_{0}^{+\infty}\di x \frac{\sqrt{x}}{(1+x) \left[\e^{c(1+x)}-1\right]}
\approx \frac{2\sqrt{2}}{\pi R^{\frac{3}{2}}} \frac{2 T_q}{T_s \sqrt{z}}\int_{0}^{+\infty}\di x \frac{\sqrt{x}}{(1+x)^2} = \frac{2\sqrt{2}}{R^{\frac{3}{2}}}\frac{T_q}{T_s} z^{-\frac{1}{2}}\ .
\end{equation}
Therefore we find the scaling $z\propto T_q^2$. In the first passage we assumed that $c\to 0$, \ie~that $z$ vanishes faster than $T_q$. If this were not the case, the expression would be at most of order $T_q^{{1}/{2}}$.

At the transition $T_s=T_c$ the finite part of~\eqref{eq:p2-zscaling-integral} vanishes, and the integral must be of order $\sqrt{z}$.
This is indeed the case if $c$ has a finite value in the $T_q\to 0$ limit, implying that $z \propto T_q$.

\subsection{Specific heat and entropy}
The energy density is
\begin{equation}\label{eq:p2-energy-int}
\varepsilon =\frac{T_s}{2} \int \di \rho(\mu)\frac{\lambda-\mu}{\e^{\beta_q T_s (\lambda-\mu)/2}-1}
= \frac{T_s}{\pi R^2} \left(\frac{z}{c}\right)^{\frac{5}{2}}
\int_{c}^{c(1+R/z)} \di y y^{\frac{3}{2}} \frac{\sqrt{2R - \frac{z}{c} y}}{\e^y-1} .
\end{equation}

In the critical regime $T_s\leq T_c$ the integral on the right hand side is of order one. 	Therefore the energy scales as $(z/c)^\frac{5}{2}=T_q^\frac{5}{2}$, and the specific heat scales as $T_q^\frac{3}{2}$.
Note that while the scaling of $z$ is different at and below $T_c$, the scaling of the energy and specific heat are the same.

Above the transition, $z$ is finite and $c\propto 1/T_q$ diverges.
The integrand in~\eqref{eq:p2-energy-int} is bounded uniformly by the exponentially large denominator, and the specific heat vanishes exponentially $\propto \e^{-c}$.

The free energy is given by
\begin{align}\nonumber
    -\beta_q f = & - \int \di \rho(\mu) \ln\left(1-\e^{-\beta_q T_s \frac{\lambda-\mu}{2}}\right)\\
    \label{eq:p2-f-int}
    = & -\frac{8}{\pi R^{3/2}} \left(\frac{T_q}{T_s}\right)^{3/2}
    \int_0^\infty \di y \sqrt{y}\ln\left(1-\e^{-c-y}\right) +\mathcal{O}(T_q^{5/2})\\
    = & \frac{4}{\sqrt{\pi R^3}} \zeta\left(\frac{5}{2}\right)\left(\frac{T_q}{T_s}\right)^{3/2}  +\mathcal{O}(T_q^{5/2})\ .
\end{align}
The free energy also vanishes as $T_q^{5/2}$. Here we used the fact that $c\to 0$, therefore the result is valid for $T_s<T_c$. However, the integral in~\eqref{eq:p2-f-int} is of order one for finite $c$, and the scaling is the same at $T_s=T_c$.

The classical $p=2$ model does not have a complex energy landscape. Therefore, we expect the entropy of the quantum model to be zero at $T_q=0$.
This is indeed the case, since $s= \beta_q (\varepsilon-f) \propto T_q^{3/2}$.

\subsection{Correlation function}
	At finite $T_q$ there is no condensation, and the equilibrium correlation function
	is
	\begin{equation}\label{eq:p2-correlator}
	C(t) =
	C_0(t) + 2 T_s \int_{0}^{2R} \di x \frac{\rho(R-x)}{z+x} \frac{\cos\left(\frac{z+x}{2} t\right)}
	{\e^{\beta_q T_s (z+x)/2}-1} \equiv C_0(t)+C_1(t)\ 
	\end{equation}
	where $C_0(t)$ is the decaying part of the ground state ($T_q=0$) correlation function, Eq.~\eqref{eq:p2-correlator-t0}.
	\begin{itemize}
		\item For $T_s>T_c$ the gap survives to $T_q\to0$. The asymptotic behaviour is the same as Eq.~\eqref{eq:p2-corr-above}.
		
		\item For {$T_s < T_c$}, note by comparing equations~(\ref{eq:p2-zscaling-integral},\ref{eq:p2-correlator}) that
		$C_1(0)=q$.
		With the change of variable $x' = x/z$, at low $T_q$
		\begin{equation}
		C_1(t) \approx 2 T_s \frac{\sqrt{z}}{c}\int_{0}^{\infty}\di x
		\frac{\rho(R - z x)}{(1+x)^2} \cos\left(\frac{1+x}{2} z t\right)\ .
		\end{equation}
		Since $\sqrt{z}\sim c \sim T_q$, the integral is of order one.
		Taking the $T_q\to 0$ at fixed time $t\sim\mathcal{O}(1)$, the time dependence
		disappears, and $C_1(t)=C_1(0)+\mathcal{O}(T_q^2)\approx q$.
		The timescale at which correlations decay is determined by the gap, $zt\sim\mathcal{O}(1)$, $t\propto \beta_q^2$.
		
		There is an intermediate regime $1\ll t \ll \beta_q^2$ in which the system
		approaches the constant value $q$, with a $t^{-\frac{1}{2}}$ power-law decay
		(given by $C_0$). At $t\propto \beta_q^2$ the correlator decays from the
		plateau to zero.
		
		\item At the transition $T_s=T_c$ the situation is similar to $T_s<T_c$, but there is no plateau ($q=0$),
		and the different scaling of $z$ implies that the timescale at which the power law is cut off
		is $t\propto \beta_q$.
	\end{itemize}

\section{Zero-temperature and harmonic approximation}
\label{app:p3}
\subsection{Scaling above the threshold at \texorpdfstring{$T_s=0$}{ts0}}
\label{app:p3-T0}
We analyse Eq.~\eqref{eq:p3-entropy-t0} in the double scaling limit with $\mathcal{E}-\ethr = A t^{*-\alpha}$ as $t^*\to+\infty$, with constant $A$ and $\alpha>0$. To ease the notation, we drop the $*$ and denote the time by $t$.
We analyse separately three terms contributing to the entropy: $I_0$ (first row), $I_1$ and $I_2$ (first and second integral, respectively).
\begin{itemize}
	\item The leading contribution to the first term is the total number of saddles at energy density $\mathcal{E}$,
	\begin{align}\nonumber
	I_0 = & \frac{1}{2}\left(1+\ln\frac{p}{2}\right)-\ethr^2+\ln|\ethr| - \left(\mathcal{E}^2-\ethr^2\right)
	+ \frac{1}{2}\re\left[\left(\frac{\mathcal{E}\mp\sqrt{\mathcal{E}^2-\ethr^2}}{\ethr}\right)^2\right]\\
	=& s_0 - 2 A \left(\ethr-\ethr^{-1}\right) t^{-\alpha} + \mathcal{O}(t^{-2\alpha})\ .
	\label{eq:t0-I0}
	\end{align}
	
	\item The leading contribution to the first integral in~\eqref{eq:p3-entropy-t0} comes from $|\omega|t \lesssim 1$.
    The edge of the semicircle is at $-p A t^{-\alpha}$.
	If $\alpha<1$, the contributing region is far from the edge. Up to exponentially small corrections
	\begin{equation}\label{eq:p3-entropy-t0-i1}
	\begin{aligned}
	-I_1 = & \int_{-c/t}^{c/t} \di \omega \rho_p(\omega+p\mathcal{E}) \ln\left(1-e^{-t|\omega|}\right)
	= \int_{-c}^{c} \frac{\di y}{t} \rho_p(p\mathcal{E}+y/t) \ln\left(1-\e^{-|y|}\right)\\
	= & 2 t^{-1} \rho_p(p\mathcal{E}) \int_0^c \ln\left(1-\e^{-y}\right)\di y + \mathcal{O}(t^{-2})
	\approx -\frac{\pi^2 p}{p-1}\sqrt{-2A\ethr} t^{-1-\alpha/2} \ .
	\end{aligned}
	\end{equation}
	To go from the first to the second line, we used that
$\rho_p(p\mathcal{E}+y/t)\approx\sqrt{A t^{-\alpha}+y t^{-1}}\approx \sqrt{A}t^{-\alpha/2}$ since $\alpha<1$.
	If $\alpha>1$, the $t^{-1}$ term dominates, and overall $I_1\propto t^{-1-\min(1,\alpha)/2}$.
	
	\item The second integral is
	\begin{equation}
	\begin{aligned}\label{eq:p3-entropy-t0-i2}
	I_2 =& t\frac{2}{\pi p^2 \ethr^2}\int_{-p(\mathcal{E}+\ethr)}^{0} \sqrt{p^2\ethr^2-(p\mathcal{E}+\omega)^2}\omega\di\omega\\
	= & -\frac{pt}{6\pi\ethr^2}\left[2(\mathcal{E}^2+2\ethr^2)\sqrt{\ethr^2-\mathcal{E}^2}
	+6\mathcal{E}\ethr^2\left(\frac{\pi}{2}+\arctan\frac{\mathcal{E}}{\sqrt{\ethr^2-\mathcal{E}^2}}\right)\right]\\
	\approx & -\frac{8p}{15\pi\ethr^2}\sqrt{-2\ethr} A^{-\frac{5}{2}} t^{1-\frac{5}{2}\alpha}~.
	\end{aligned}
	\end{equation}
\end{itemize}

Summing the three contributions, the entropy of stable states at $t$, $\mathcal{E}-\ethr = A t^{-\alpha}$ is given by
\begin{equation}
s-s_0 = c_0 t^{-\alpha} - c_1 t^{-1-\min(\alpha,1)/2} - c_2 t^{1-\frac{5}{2}\alpha}~,
\end{equation}
where the $c$'s are positive coefficients depending on $A$ and $p$.

As expected from the intuition given in the main text, there is a competition between the positive contribution from the first term, and the negative ones from the other two.
The scaling of the maximum is obtained by requiring that the first and third term have the same exponent (the second is subleading), fixing to $\alpha = 2/3$.
Maximising the coefficient $c_0-c_2$ fixes $A$, leading to the scaling,
\begin{equation}\label{eq:t0-smax}
s_M(t) = s_0 + c_M t^{-\frac{2}{3}}~, \qquad c_M = \frac{3(3\pi)^\frac{2}{3}(p-2)^{5/3}}{5\cdot 2^\frac{1}{3}(p-1)^\frac{1}{3} p^2}.
\end{equation}

\subsection{Harmonic approximation}
At low $T_s$ and for short enough times, the classical dynamics of the $p$-spin model can be approximated by expanding the potential to second order around each stationary point of the energy landscape.

With the change of basis~\eqref{eq:FP-to-H}, the Fokker--Planck operator is mapped to the Hamiltonian of quantum harmonic oscillators of frequencies
\begin{equation}\label{eq:harmonic-omega}
  \omega_\mu^2 = \hat{\lambda}T_s + (\lambda-\mu)^2 .
\end{equation}
The $\mu$'s are the eigenvalue of the Hessian, and are distributed with a semicircle law of radius $R=\sqrt{2p(p-1)}$~\cite{Cavagna1998}.
As discussed in Ref.~\cite{Kurchan09-lectures}, and noting the role of $\hat{\lambda}$, the spectrum of $H$ is given by
\begin{equation}\label{eq:harmonic-spectrum}
E_n^{(\mu)}[\lambda,\hat{\lambda}] = T_s\omega_\mu
\left(n+\frac{1}{2}\right)-\frac{T_s}{4} (\lambda-\mu)\ ,
\end{equation}
and the contribution to the partition function from each mode is 
\begin{equation}
Z^{(\mu)}[\lambda,\hat{\lambda}]
=
\frac{\e^{t^*[\hat{\lambda}/2-\omega_{\mu}+(\lambda-\mu)/2]}}{1-\e^{-2t^*\omega_\mu}}~,
\end{equation}
where $t^*=\beta_q T_s/2$. Note that both stable and unstable classical degrees of freedom are mapped to quantum harmonic oscillators, with a spectrum shifted as in~\eqref{eq:harmonic-spectrum}.

The total partition function is obtained by the maximisation
\begin{equation}\label{eq:harmonic-pf}
	\ln Z = \max_{\mathcal{E}, \lambda, \hat{\lambda}}
	\left\{I_0(\mathcal{E})
	- \int \di\rho(\mu) \left[\ln\left(1-\e^{-2t^*\omega_\mu}\right)+
	t^*\left(\omega_\mu-\frac{\lambda-\mu}{2}\right)\right]
	+ \frac{\hat{\lambda} t^*}{2}\right\}\ ,
\end{equation}
over the energy and the two Lagrange multipliers.
Maximising over $\lambda$ fixes $\lambda=-p\mathcal{E}+\mathcal{O}(T_s)$~\cite{BiroliKurchan01}.
In the following we show that the spherical constraint fixes the relative scaling of $\hat{\lambda}$ and $\lambda$, and that maximising over $\mathcal{E}$ ultimately gives the same scaling above the threshold as in the $T_s=0$ case.

\subsubsection{Spherical constraint}
\begin{figure}
	\centering
	\includegraphics[width=.5\textwidth]{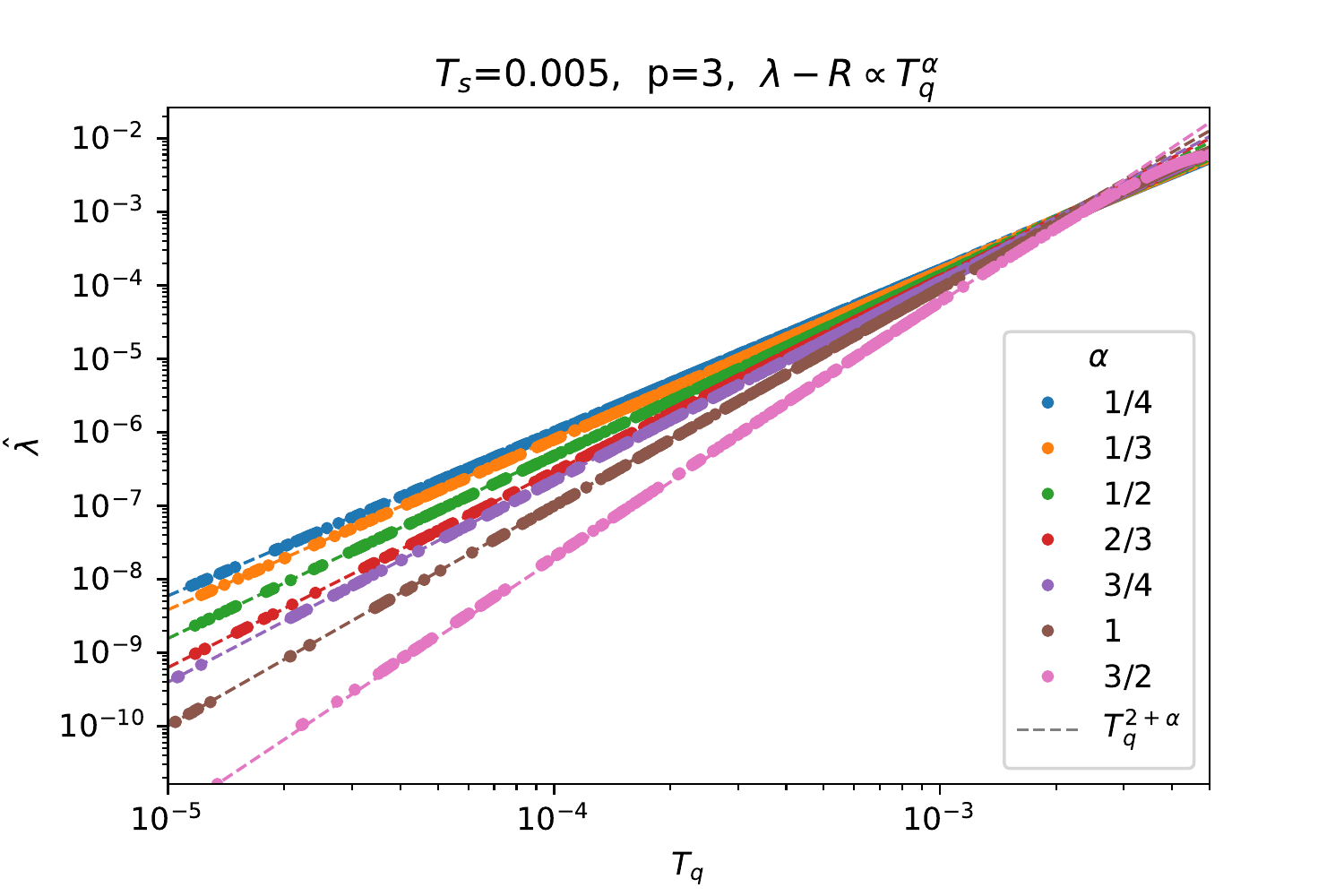}
	\caption[Spherical constraint in the harmonic approximation]{Scaling of $\hat{\lambda}$ at $T_q\to 0$ in the harmonic approximation. From the numerical
		solution of~\eqref{eq:spherical-const_Ts} with
		$\lambda-R\propto T_q^\alpha$. For each $\alpha$, the dashed line shows the $T_q^\beta$, $\beta=\alpha+2$ scaling.}\label{fig:lhalphascaling}
\end{figure}

Maximising~\eqref{eq:harmonic-pf} over $\hat{\lambda}$ leads to the spherical constraint
\begin{equation}\label{eq:spherical-const_Ts}
\int \di\mu \frac{\rho(\mu)}{2\omega_\mu} \coth\left(\frac{\beta_q T_s}{2} \omega_\mu \right) = \frac{1}{T_s}~,\
\end{equation}
which has a form similar to the $p=2$ case~\eqref{eq:Fint1}, but with $\omega_{\mu}$ given by the harmonic approximation relation~\eqref{eq:harmonic-omega}.
To fix the relative scaling of the two Lagrange multipliers $\lambda, \hat{\lambda}$, we consider the $T_q\to 0$ limit with $\lambda-R\propto T_q^\alpha$, $\hat{\lambda}\propto T_q^\beta$.
Solving~\eqref{eq:spherical-const_Ts} numerically we find that $\beta=2+\alpha$, see Fig.~\ref{fig:lhalphascaling}.

\subsubsection{Scaling above the threshold}
We now study the scaling above the threshold of~\eqref{eq:harmonic-pf}, comparing it with the analysis of the $T_s=0$ case (Sec.~\ref{app:p3-T0}).
\begin{itemize}
	\item The first term $I_0(\mathcal{E})$ is exactly the same, counting the number of stable states at energy $\mathcal{E}$.
	
	\item The first integral corresponds to $I_1$~\eqref{eq:p3-entropy-t0-i1}.
	Note that since $\omega>|\lambda-\mu|/2$,
	the $T_s>0$ contribution
	is smaller than $I_1$, which was shown to be always subleading in the 
	previous section,
	
	\begin{equation}\begin{split}
	\left|\int \di\rho(\mu) \ln\left(1-\e^{-2t^*\omega_\mu}\right)\right|
	= -\int \di\rho(\mu) \ln{\left|1-\e^{-2t^*\omega_\mu}\right|}
	\\
	< -\int \di\rho(\mu) \ln{\left|1-\e^{-t^*|\lambda-\mu|}\right|} = |I_1| \ .
	\end{split}
	\end{equation}
	
	\item For $\hat{\lambda}T_s\to 0$,
	the second integral reduces to $I_2$~\eqref{eq:p3-entropy-t0-i2}.
	The correction can be separate into two contributions.
	The contribution from $|\lambda-\mu|\lesssim T_q^{\beta/2}$ is bounded by $\approx T_q^{\beta+\alpha/2} = T_q^{2+3\alpha/2}$. Expanding the
	square root for $|\lambda-\mu|\gg T_q^{\beta/2}$,
	\begin{equation}\label{eq:i2-correction}
	\approx\frac{1}{T_q}\int_{|\lambda-\mu|\gg T_q^\beta} \di\rho(\mu) \frac{T_q^\beta}{|\lambda-\mu|}
	\propto T_q^{1+\frac{3}{2}\alpha} \ln(T_q)
	\end{equation}
	and we get a logarithmic correction,
	which is small compared to $I_2\propto T_q^{\frac{5}{2}\alpha-1}$.
	\item The additional term is $\frac{\hat{\lambda} T_s}{4T_q} \propto T_q^{1+\alpha}$, and is always subleading.
\end{itemize}

Therefore, the scaling~\eqref{eq:t0-smax} is unchanged within the harmonic approximation and $\alpha=2/3$.

\subsection{Correlation functions}
Analysing the model from the classical and quantum points of view leads to two equivalent constructions for the path integral (MSRJD and Matsubara, respectively).
Both path integrals are expressed in terms of correlation functions, which are different from each other, but related by the change of basis~\eqref{eq:op-change}.
Note that the function $C$ is the same in both basis, while $R$ and $D$ change by terms that vanish in the $T_s\to 0$ limit.
Within the harmonic approximation, it is simpler to work directly on the quantum side, calculating correlation
functions in terms of harmonic oscillators, as for $p=2$~\eqref{eq:p2-correlator}.
Real-time correlation functions are obtained as Fourier integrals involving the density of states~\eqref{eq:p3-corr} with the density of states given in Eq.~\eqref{eq:deformed-DoS} and Fig.~\ref{fig:deformed-sc}.

\bibliography{syk-glass.bib}

\end{document}